\documentclass[preprint,prd,preprintnumbers,amsmath,amssymb,nofootinbib,eqsecnum]{revtex4}

\usepackage{graphicx}
\usepackage{dcolumn}
\usepackage{bm}
\newcommand{\Mat}[1]{{{\boldsymbol{#1}}}}
\newcommand{\abs}[1]{\left\vert#1\right\vert}
\def\be{\begin{equation}}
\def\ee{\end{equation}}
\def\dd{\mathrm{d}}

\begin{document}

\title{\bf Equations of motion according to the asymptotic post-Newtonian scheme for general relativity in the harmonic gauge}

\author{Mayeul Arminjon}
\affiliation{Laboratoire ``Sols, Solides, Structures'' \\
(CNRS / Universit\'e Joseph Fourier / Institut National Polytechnique de Grenoble) 
\\ BP 53, F-38041 Grenoble cedex 9, France. \\
Email: arminjon@hmg.inpg.fr\\
[Currently at Dipartimento di Fisica, Universit\`a di Bari,
Via Amendola 173, I-70126 Bari, Italy.]}

\date{September 14, 2005}
\begin{abstract}
The asymptotic scheme of post-Newtonian approximation defined for general relativity (GR) in the harmonic gauge by Futamase \& Schutz (1983) is based on a family of initial data for the matter fields of a perfect fluid and for the initial metric, defining a family of weakly self-gravitating systems. We show that Weinberg's (1972) expansion of the metric and his general expansion of the energy-momentum tensor ${\bf T}$, as well as his expanded equations for the gravitational field and his general form of the expanded dynamical equations, apply naturally to this family. Then, following the asymptotic scheme, we derive the explicit form of the expansion of ${\bf T}$ for a perfect fluid, and the expanded fluid-dynamical equations. (These differ from those written by Weinberg.) By integrating these equations in the domain occupied by a body, we obtain a general form of the translational equations of motion for a 1PN perfect-fluid system in GR. To put them into a tractable form, we use an asymptotic framework for the separation parameter $\eta $, by defining a family of well-separated 1PN systems. We calculate all terms in the equations of motion up to the order $\eta ^3$ included. To calculate the 1PN correction part, we assume that the Newtonian motion of each body is a rigid one, and that the family is quasi-spherical, in the sense that in all bodies the inertia tensor comes close to being spherical as $\eta \rightarrow 0$. Apart from corrections that cancel for exact spherical symmetry,  there is in the final equations of motion one additional term, as compared with the Lorentz-Droste (Einstein-Infeld-Hoffmann) acceleration. This term depends on the spin of the body and on its internal structure. 
\end{abstract}

\maketitle

\section{Introduction}

Explicit, tractable equations of motion of celestial bodies can be used to compute ephemerides which have to represent the prediction of a given theory of gravitation for these motions. Such tractable, actually-used equations of motion are among the most important ones in a theory, of course. Naturally also, such equations are necessarily of an approximate nature, and this is all the more so in relativistic theories of gravitation, all of which are much more complex than is Newton's theory. Most of the parameters that enter these equations of motion are adjusted for the very construction of the ephemerides, or at least are adjusted under the assumption that the theory considered is correct \cite{Newhall83,Muller92}. Therefore, the ephemerides represent in fact a  fitting of astronomical observations by the given theory of gravitation or, more precisely, by the actually-used, approximate equations of motion derived from that theory. Einstein's general relativity (GR) is by very far the mostly investigated theory, and the literature about the post-Newtonian (PN) equations of motion for the mass centers (EMMC's) of an isolated system of celestial bodies in GR is quite vast. However, the EMMC's that are actually used in relativistic celestial mechanics \cite{Newhall83,Muller92,Moisson01} were obtained two years only after the proposal of GR: these are the Lorentz-Droste (LD) equations \cite{Lorentz-Droste}. They are more widely known as the Einstein-Infeld-Hoffmann equations, because that work of Lorentz and Droste \cite{Lorentz-Droste} seems to have been forgotten until rather recently \cite{DSX91}, and because these same equations were derived (though under different assumptions and for two bodies only) by Einstein and his coworkers \cite{EIH}. \\

The most satisfactory derivation of the LD equations is that provided by Damour, Soffel and Xu (hereafter ``DSX" for short) \cite{DSX91}. As shown by DSX, the LD equations are the ``monopole-truncated" EMMC's, corresponding to the ideal case where the gravitational field of each body is characterized by just one parameter (its constant mass), all higher-order multipole moments and all spin moments being zero. For each body, a set of such moments is defined by these authors \cite{DSX91}: generalizing results obtained by Blanchet and Damour (which were based on an exact multipole expansion of the retarded potential due to a compact source) \cite{BlanchetDamour89}, these moments are defined from multipole integrals, limited to this body, of the relevant current and energy densities; these moments then determine the coefficients in multipole series expansions of the (external) PN gravitational potentials produced by this body. The results of DSX \cite{DSX91,DSX92}, in particular the exact recovering of the LD equations as the monopole-truncated EMMC's, are further confirmed by Racine and Flanagan \cite{RacineFlanagan} who also consider multipole expansions, but use a somewhat different approximation method (based on a separation between Newtonian and (first) post-Newtonian equations and, therefore, leading to Poisson equations instead of d'Alembert equations for the gravitational potentials; moreover, in order to extend the validity of the results to strong fields, Racine and Flanagan \cite{RacineFlanagan} use surface-integrals definitions of the multipole moments and the mass centers). A question arises as to the relevance of the monopole-truncated model (or, shortly, the monopole model) leading to the LD equations: does this model approach the relevant exact general-relativistic motion accurately enough? (The fact that the LD equations do allow to compute accurate ephemerides does not give a strong argument for answering ``yes:" from the purely logical point of view, one cannot preclude that the observational agreement could be less good if one would use a more accurate approximation.) Within the DSX framework \cite{DSX91}, the answer to this question should be obtained by incorporating higher-order multipole moments and spin moments, and by checking the magnitude of the corresponding new terms in the equations of motion. DSX \cite{DSX92} have indeed studied a ``monopole-dipole" model for which, in addition to the monopole mass moment, also the spin dipole moments are non-zero, and in fact may be considered constant. Although DSX \cite{DSX92} did not discuss the magnitude of the new (spin-orbit and spin-spin) terms in the equations of motion, one may guess that they should be very small in the solar system, because they turn out to be of a high order in the separation parameter. However, in view of the infinite number of multipole and spin moments, there is no a priori guarantee that the set of the other moments gives a negligible contribution. Indeed DSX ``{\small state clearly that these models} [such as the monopole model and the monopole-dipole model] {\small do not need to represent the first step in some asymptotic approximation to reality, but only to be able to ``save the phenomena" with an acceptable accuracy, and in a logically consistent manner.}" \cite{DSX92}\\

By definition, neither in the monopole model nor in the monopole-dipole model does the {\it internal structure} of the gravitating bodies influence the equations of motion. However, there are indications that in fact the internal structure might play a non-negligible role. There is first a qualitative argument: in any relativistic theory, the mass-energy equivalence implies that any kind of energy should both contribute to the gravitational field and be subjected to its influence. Thus, for instance, the rest-mass energy, but also the energy due to the interaction between matter and gravitational field, of which the Newtonian potential energy gives a first approximation, should play a role---as does the kinetic energy due to the internal motion: the influence of the latter is attested by the presence of the spins in the equations of motion derived from the monopole-dipole model. And since the distribution of the energy among these different forms can vary from one celestial body to the other, one a priori expects that the different distributions might affect the motion. A more quantitative argument follows from a derivation of the EMMC's in an alternative scalar theory, which has been done recently \cite{A25,A26,A32}. Indeed, it has been found there that several structure parameters enter the explicit EMMC's tailored to eliminate numerically-negligible terms, and it has been argued that the same should occur ``in nearly any other theory" if a similar method was used \cite{A32}. \\

Therefore, the {\it goal of the present investigation} was to check whether this conjecture about the influence of structure parameters does apply {\it to GR}. The method used in Refs. \cite{A25,A26,A32}, which will also be used in the present work, is very different from that followed by DSX \cite{DSX91,DSX92} and by Racine and Flanagan \cite{RacineFlanagan}, who use multipole expansions from the beginning and who, in a first step, analyse the equations for the moments in reference frames attached to the different bodies.
\footnote{
Our method is closer to that used by Fock \cite{Fock64}, but it differs from it in that we use the ``asymptotic" scheme of PN approximation \cite{FutaSchutz,A23}, instead of the ``standard" scheme developed by him \cite{Fock64} and by Chandrasekhar \cite{Chandra65}. The difference between these two schemes will be demonstrated in Sect. \ref{AsymptoGR}, especially Subsect. \ref{Explicit-expansions}. In the particular case of a test particle in a Schwarzschild field, these two schemes are yet equivalent \cite{A29}. Another important difference is that we use also a definite asymptotic framework for the separation parameter, which is exposed in Sect. \ref{GoodSeparation}. 
}
Instead, our first step consists simply of an integration of the PN field equations for a perfect fluid in the global reference frame, say F, which provides a {\it general} (but not tractable) form of the PN EMMC's---the mass centers themselves being defined as local barycenters of the PN rest-mass density in the frame F \cite{A25}. Such general PN EMMC's were not given in previous works, except in the form involving infinite series of multipole and spin moments \cite{RacineFlanagan}, which is quite complex. Of course, our restriction to perfect fluids means some loss of generality as compared with the approach of Refs. \cite{DSX91,DSX92,RacineFlanagan}, which does not need to consider a particular material model; but this restriction is not a serious one in the solar system \cite{Will71a}, essentially because, inside massive bodies, the deviatoric stresses are small as compared with the isotropic pressure. (Moreover, the present method might be used with a more sophisticated material model, at the price of redoing the calculations.) In a second step, to get tractable equations, we use three simplifications which do occur for the solar system: {\bf i}) the fact that its main bodies are well-separated. This is an essential assumption, which is set by assuming that a precise separation parameter \cite{A26}, which we denote by $\eta$, is a small number $\eta_0$ for the system of interest, S, and by introducing a family $(\mathrm{S}^\eta)$ of well-separated systems (each system is weakly gravitating, with the field strength being nearly the same for all systems, and with $\mathrm{S}^{\eta_0} = \mathrm{S}$). It enables one to define a hierarchy of well-defined approximations for the ``tidal" effects, by calculating asymptotic expansions, with respect to $\eta $, of the integrals that enter the general form of the PN EMMC's \cite{A32}. {\bf ii}) The second simplifying feature of the solar system is that its main bodies have a nearly rigid rotation about themselves. {\bf iii}) Lastly, the main bodies of the solar system are nearly spherical. We use the sphericity assumption only for the zero-order rest-mass density, $\rho_0$ (but this implies that the zero-order pressure and the self Newtonian potential are also spherically symmetric in each body). But, as well as for the rigidity assumption, we use the sphericity assumption merely at the stage of calculating the PN {\it corrections}, not for the zero-order calculations themselves. Moreover, we shall simultaneously derive an explicit (though less condensed) form of the equations of motion {\it without} using this sphericity assumption, and using instead a much weaker ``quasi-sphericity" assumption, expressed by Eq.~(\ref{quasi-spheric}). 
\footnote{
Thus, the incompatibility between exact spherical symmetry and non-zero self-rotation is not relevant here, firstly because we do give the formulas for the non-spherical case. Secondly, when we use the assumption of spherical zero-order densities to get a more tractable expression of the PN corrections, this assumption has the status of a relevant approximation. According to this view, the zero-order calculations might take into account the departure from sphericity. The spherical zero-order density $\rho_0$ considered when calculating the PN corrections might then be defined as some relevant approximation (a least-squares approximation, say) to the exact, non-spherical zero-order density. The same can be said about the rigidity assumption.
}   
We argued previously \cite{A26} that, in the solar system, the sphericity assumption is likely to lead to relative errors smaller than $10^{-3}$ when calculating the PN corrections, which are already very small. In summary, we believe that the assumptions introduced in the present paragraph are enough justified in the solar system. The ``asymptotic" PN scheme, which is used here, leads to a separation between zero-order (Newtonian) and first-order (1PN) equations. Therefore, the small effects neglected, such as the departure from rigidity or the shear stresses \cite{XuWuSoffelKlioner03}, could be described accurately enough by taking them into account merely for the Newtonian calculations. (Such Newtonian calculations using a more general model of the bodies are, of course, a well-developed subject in geophysics.) Hence, due to this separation in the scheme used here, it would not make much practical sense to include such small effects into the calculations of the PN corrections, unless at the same time one would compute {\it second} post-Newtonian corrections to the motion.\\

This paper is organized thus: the next Section summarizes the asymptotic-expansion method of the local fields and equations for a perfect fluid in GR under the harmonic gauge. The general form of the equations of motion for a perfect-fluid system in GR (in the harmonic gauge) is derived in Sect. \ref{EMMC-GR-general}. The framework which is used to rigorously account for the good separation between bodies is summarized in Sect. \ref{GoodSeparation}. Based on Appendix \ref{Integrals-eta}, in which we compute the integrals entering the general form of the EMMC's, for the case of well-separated, rigidly-rotating and quasi-spherical bodies, Sect. \ref{EMMC-GR-explicit} presents the explicit EMMC's obtained for that case, together with their specialization to the case with spherical mass densities. Our conclusion is presented in Sect. \ref{Conclusion}.

\section{Asymptotic P.N. approximation for a perfect fluid in G.R. under the harmonic gauge} \label{AsymptoGR}

The standard PN approximation is based on the classical works of Fock \cite{Fock64} and Chandrasekhar \cite{Chandra65}. At those times, the numerical analysis of partial differential equations (PDE's) was far less developed than it is now. It would now seem relevant to build an approximation scheme in agreement with the asymptotic schemes currently used in the numerical analysis of PDE's. In a such scheme, one has a regular {\it family} of systems, $(\mathrm{S}_\epsilon)$ (each system being defined by a boundary-value problem---an initial-value problem for that matter, because relativistic gravitational equations are hyperbolic), so that one may make the ``small parameter" $\epsilon $ as small as desired, indeed. Then, for the corresponding family of fields, one should be able to state some asymptotic expansions with respect to $\epsilon$; in particular, all of the unknown fields should be expanded, because in general all depend on $\epsilon$. Moreover, the family should be associated in a physically natural way with the system of interest S, which should correspond to a small value $\epsilon _0$ of the parameter, $\mathrm{S}=\mathrm{S}_{\epsilon_0}$, so that it makes sense to use the expansions for the given system S. Finally, $\epsilon$ should be the natural field-strength parameter \cite{Fock64}, so that a small value of $\epsilon$ means indeed a weak gravitational field. It has been shown by the work of Futamase and Schutz \cite{FutaSchutz} how to develop a post-Newtonian approximation along this line for GR (in the harmonic gauge). Below, we indicate first (Subsects. \ref{Base-expansions} and \ref{Expansion-field-eqs}) how, starting from their initial condition, one indeed obtains the general expansions and expanded equations derived by Weinberg in Sections 9.1 and 9.3 of his classical book \cite{Weinberg}. Our aim is not to present mathematically rigorous proofs but to show that the main features of Weinberg's approach occur rather naturally and convincingly from the point of view of asymptotic analysis. Futamase and Schutz \cite{FutaSchutz} did not investigate the relation between their ``asymptotic" PN approximation and Weinberg's expansions, which, in our opinion, remained difficult to understand. Then we show (Subsect. \ref{Explicit-expansions}) that, however, the explicit expansion of the energy-momentum tensor ${\bf T}$ and the explicit expanded equations obtained according to this ``asymptotic scheme" for a perfect fluid are different from the equations stated by Weinberg in his Section 9.8.

\subsection{Expansion of the matter fields and the metric} \label{Base-expansions}
Futamase and Schutz \cite{FutaSchutz} assume the following initial data for the fields $p$ (pressure), $ \mu^{\ast}$ (proper energy density in {\it mass density} units),
\footnote{ 
Futamase and Schutz \cite{FutaSchutz}, as well as Weinberg \cite{Weinberg}, use the notation $\rho $ for $\mu ^{\ast}$. Moreover, they set $c=1$, so that the physical dimensions are not apparent in their equations.
}  $\mathbf{u}$ (coordinate velocity), and $\Mat{g}$ [space-time metric; in this paper, we follow the conventions of Weinberg \cite{Weinberg}, in particular the signature is (3,1)]: at time $t=0$,
\begin{equation}\label{PN_IC1}
p^{(\epsilon)}(\mathbf{x})=\epsilon^4 p^{(1)}(\mathbf{x}),
\ee 
\begin{equation}\label{PN_IC2}
\mu^{\ast (\epsilon)}(\mathbf{x})=\epsilon^2 \mu^{\ast (1)}(\mathbf{x}),
\end{equation}
\begin{equation}\label{PN_IC3}
\mathbf{u}^{(\epsilon)}(\mathbf{x})= \epsilon\, \mathbf{u}^{(1)}(\mathbf{x}),
\end {equation}
\be \label{ICmetric-F&S}
\sqrt{- g^{(\epsilon)}}\ g^{(\epsilon)\,ij} ({\bf x}) = \delta _{ij}, \qquad
\left(\sqrt{-g^{(\epsilon)}}\ g^{(\epsilon)\,ij} \right)_{,0} ({\bf x}) =0 \qquad (1\leq i\leq 3,\quad 1\leq j\leq 3)
\ee
[here $g \equiv \text{det}(g_{\mu \nu })$, $(g^{\mu \nu })$ is the inverse matrix of $(g_{\mu \nu })$, and ${\bf x} \equiv (x^i)$ is the spatial position]. Conditions (\ref{PN_IC1})-(\ref{PN_IC3}) for the matter fields are suggested by an exact similarity transformation valid for the Euler-Newton equations, which defines the weak-field limit for a perfect-fluid system in Newton's theory \cite{FutaSchutz,A23}. This indeed suggests to define the initial conditions for the Newtonian limit of any ``relativistic" theory by applying the similarity transformation to the initial data, at least for the matter fields---and this leads exactly to Eqs.~(\ref{PN_IC1})-(\ref{PN_IC3}). As to condition (\ref{ICmetric-F&S}), it is just the condition that one would impose if one would wish to have the following ``conformally-Euclidean" (or isotropic) form for the space metric $\Mat{\gamma }$ (the spatial part of $\Mat{g}$ in the frame defined by the coordinate system):
\be \label{spatial-F&S}
\gamma _{ij} = \sqrt{-g}\ \delta _{ij}, 
\ee
without imposing any a priori restriction on the factor $\sqrt{-g}$. (Hence, it is legitimate to postulate this condition (\ref{ICmetric-F&S}), but it is not clear that one {\it has to }postulate it, unless one does so precisely to enforce spatial isotropy.) Note that condition (\ref{ICmetric-F&S}) does not depend on $\epsilon $. Thus, using first Eqs.~(\ref{PN_IC1})-(\ref{PN_IC3}) to go from the finite small value $\epsilon _0$, valid for the physically given system S, to $\epsilon =1$, and then once more to go from $\epsilon =1$ to the arbitrary value $\epsilon $, one indeed naturally associates with S a family $(\mathrm{S}_\epsilon)$ of systems. \\

Now, let us change the mass and time units for system $\mathrm{S}_\epsilon$ in this way \cite{A23}: $[\mathrm{M}]_\epsilon = \epsilon^2 [\mathrm{M}]$ and $\ [\mathrm{T}]_\epsilon = [\mathrm{T}]/\epsilon$, where $[\mathrm{M}]$ and $[\mathrm{T}]$ are the units for system S$_1$. In these units, the initial data~(\ref{ICmetric-F&S}) for the metric is unchanged (because the metric is adimensional), but Eqs.~(\ref{PN_IC1})-(\ref{PN_IC3}) become simply 
\begin{equation}\label{PN_IC1-3-varying}
p^{(\epsilon)}(\mathbf{x})=p^{(1)}(\mathbf{x}),
\quad \mu^{\ast (\epsilon)}(\mathbf{x})= \mu^{\ast (1)}(\mathbf{x}),
\quad \mathbf{u}^{(\epsilon)}(\mathbf{x})= \mathbf{u}^{(1)}(\mathbf{x}).
\end {equation}
Thus, the initial condition is {\it independent of} $\epsilon$, moreover we have just $\epsilon =c^{-1}$ (we take the velocity of light to be $c=1$ {\it in the units for system} S$_1$). The initial-value problem still depends on $\epsilon$ {\it or more exactly on $\epsilon^2$,} since in general units the Einstein equations involve the square $c^2=\epsilon^{-2}$:
\be\label{HilbertEinstein}
R^{\mu \nu} -\frac{1}{2}g^{\mu \nu} R^\lambda _\lambda  = -\frac{8\pi G}{c^2} T^{\mu \nu}
\ee
(this writing is in accordance with Weinberg's, which itself follows his conventions about the Riemann and Ricci tensors; we take ${\bf T}$ in units of mass density, thus ML$^{-3}$; note that $G$ with dimension L$^3$M$^{-1}$T$^{-2}$ is invariant in the change of units). The effective small parameter is thus $\epsilon^{2}$, rather than $\epsilon$. Hence, it suggests itself to state Taylor expansions with respect to $\epsilon^2=c^{-2}$, starting from the {\it zero-order} term (denoting with a prime the fields as expressed in these varying units):
\be 
p' = p'_0 + p'_1 c^{-2}+ O(c^{-4}),\qquad \mu'^{\ast}  = \mu'^{\ast}_0 + \mu'^{\ast} _1 c^{-2}+ O(c^{-4}),\qquad {\bf u}' = {\bf u}'_0 + {\bf u}'_1 c^{-2}+ O(c^{-4}), 
\ee
etc. When coming back to the fixed units (of system S$_1$), these expansions are modified in a straightforward way: 
\be \label{matter-expansion}
p = [p_0 + p_1 \epsilon^2+ O(\epsilon^4)]\epsilon^4,\qquad \mu^{\ast}  = [\mu^{\ast} _0 + \mu^{\ast} _1 \epsilon^2+ O(\epsilon^4)]\epsilon^2 ,\qquad {\bf u} = [{\bf u}_0 + {\bf u}_1 \epsilon^2+ O(\epsilon^4)]\epsilon. 
\ee
On the other hand, the definition of tensor ${\bf T}$ from the matter fields is for a perfect fluid
\be \label{T-fluid}
T^{\mu \nu } = (\mu^{\ast} +p c^{-2}) U^\mu U^\nu + p c^{-2} g^{\mu \nu }
\ee
\cite{Fock64,Weinberg}, where $U^\mu$ is the four-velocity of the fluid:
\be\label{4-velocity}
U^\mu  \equiv \dd x^\mu /\dd s, \quad \dd s \equiv (-g_{\mu \nu }\dd x^\mu \dd x^\nu )^{1/2},
\ee
hence 
\be\label{U^i}
U^i=U^0u^i/c
\ee
since $u^i\equiv \dd x^i/\dd t$ and $x^0=ct$. Therefore, the definition of ${\bf T}$ from the matter fields does involve directly $c^{-1}=\epsilon$. However, the  equation $T^{\mu \nu } _{;\nu } =0$, once written in terms of the matter fields using this definition, will contain only $c^{-2}$. (This is not immediate to check on the usual forms of the exact perfect-fluid equations in GR, e.g.~Font \cite{Font}, because usually it is set $c=1$; see Ref. \cite{A23}, Eqs.~(3.8) and (3.10), for the equations in an alternative theory, based on a slightly different conservation equation.) Thus, this passage by the $\epsilon $-dependent units at least suggests that the expansions in the usual field-strength parameter $\epsilon$ \cite{Fock64,Weinberg,MTW,Will93}, or the formal expansions in $1/c$ \cite{Chandra65}, could be in fact expansions in the {\it square} $\lambda \equiv \epsilon ^2$, or $1/c^2$. (One sometimes invokes the invariance by time-reversal, but this applies to the equations, not to their solutions for a generic situation.) As stated by Weinberg \cite{Weinberg}, ``the real justification for these expansions will come below when we show that they lead to a consistent solution of the Einstein equations." We will use below the varying units for the metric, whose expansion is less straightforward. \\

In order that there are well-defined zero-order fields and equations, it is necessary that the metric has a finite limit as $\epsilon \rightarrow 0$. This should occur in the fixed units, for we know that in a real weak field, thus for a finite small value $\epsilon _0$ of the field strength, we can approximate the metric by a finite metric which is independent of $\epsilon _0$, namely a flat one. Thus, without anticipating the flatness, the metric should have an expansion with a first term of order zero:
\be \label{0order-metric-expansion}
\Mat{g} = \Mat{g}_0 + O(\epsilon ^{2}). 
\ee 
Setting  $\Mat{\eta }\equiv \mathrm{diag}(-1,1,1,1)$ and $\Mat{h}\equiv \Mat{\eta} -\sqrt{-g}\, \Mat{g}$, the harmonic gauge condition writes $h^{\mu \nu }_{,\nu }=0$ and in this gauge the Einstein equations take the (``relaxed") form \cite{FutaSchutz}
\be
\square h^{\alpha \beta } \equiv -\eta ^{\mu \nu }h^{\alpha \beta } _{,\mu \nu }=16\pi \Lambda^{\alpha \beta },
\ee
where $\Lambda^{\alpha \beta }$ is the sum of a term linear in tensor ${\bf T}$ and a polynomial $ P^{\alpha \beta }(h^{\mu \nu },h^{\mu \nu}_{,\rho },h^{\mu \nu}_{,\rho \sigma }) $, having at least quadratic terms. The zero-order equations are hence simply
\be \label{0-order-h}
\square h_0^{\alpha \beta } =16\pi \Lambda_0^{\alpha \beta }
\ee 
(with $h_0^{\alpha \beta }$ and $\Lambda_0^{\alpha \beta }$ the zero-order coefficients in the expansions of $h^{\alpha \beta }$ and $\Lambda^{\alpha \beta }$). From the expansions (\ref{matter-expansion}), it results that tensor ${\bf T}$ is $O(\epsilon^2)$ (in the fixed units) and hence makes no contribution to $\Lambda _0^{\alpha \beta } $. Hence, we have $\Lambda_0^{\alpha \beta }=P^{\alpha \beta }(h_0^{\mu \nu },h^{\mu \nu}_{0,\rho},h^{\mu \nu}_{0,\rho \sigma} ) $. Thus, the zero-order equations (\ref{0-order-h}) are still {\it nonlinear}, just as the equations for the exact field $\Mat{h}$. However, $\Mat{h}_0={\bf 0}$  (hence $\Mat{g}_0 = \Mat{\eta }$) is a solution of these equations which is compatible with the initial data deduced from (\ref{ICmetric-F&S}), $\Mat{h}(t=0,{\bf x})={\bf 0}$ and $\partial_0\Mat{h}(t=0,{\bf x})={\bf 0}.$
Moreover the solution of the initial-value problem for these nonlinear wave equations should be unique. Thus the limiting metric is flat indeed:
\be \label{0-order-metric}
\Mat{g}_0 = \Mat{\eta }. 
\ee 
However, if one reexpresses this limit in our varying units (or, for that matter, if one just uses the varying unit of time, thus replacing the time $t$ by the ``dynamical time" $t'\equiv \epsilon t$ \cite{FutaSchutz,RacineFlanagan}), he finds that the limit $\Mat{g}_0$ has the following components, one of which is singular at the limit $\epsilon \rightarrow 0$:
\be \label{g0-varying}
(_0g'_{\mu \nu })= \mathrm{diag}(-\epsilon ^{-2},1,1,1) = \mathrm{diag}(-c^2,1,1,1).
\ee
From (\ref{g0-varying}), it follows that, to 1PN order, i.e., including terms up to the order $c^{-2}$ in the varying units, it is natural to postulate the following expansions:
\be \label{g00-expansion-c2}
g'_{00 } = -c^2  + \,_1g'_{00 } +\, _2g'_{00 }\,c^{-2}    + O(c^{-4}), 
\ee 
\be \label{gij-expansion-c2}
g'_{ij } =  \delta _{ij }  +\, _1g'_{ij }\,c^{-2}  + O(c^{-4}), 
\ee 
\be \label{g0i-expansion-c2}
g'_{0i } = \,_1g'_{0i }\,c^{-2}  + O(c^{-4}).
\ee 
This leads, in the fixed starting units, to 
\be \label{metric-expansion-fixed}
g_{00 } = -1+ \,_1 g_{00 }\,\epsilon ^2 +\,_2 g_{00 }\,\epsilon ^4   + O(\epsilon ^6), \quad g_{ij } =  \delta _{ij }  +\, _1g_{ij }\,\epsilon ^{2} + O(\epsilon ^4), \quad g_{0i } =\,  _1g_{0i }\,\epsilon ^{3} + O(\epsilon ^5).
\ee 
This is the expansion written by Weinberg \cite{Weinberg} (accounting for the fact that this author incorporates the small parameter in the expansion coefficients) and used by Racine and Flanagan \cite{RacineFlanagan}.
\\

The asymptotic expansions of the fields have first of all to be valid at fixed values of the time and space variables, of course. Since the velocities in the system $\mathrm{S}_\epsilon $ vary like $\epsilon $, it follows that the characteristic times (e.g. the orbital periods) vary like $ \epsilon ^{-1}$. Therefore, the relevant time variable in the expansions is not the time $t$ in fixed units, but the ``dynamical time" $t'\equiv \epsilon t$ \cite{FutaSchutz,A23,RacineFlanagan}. This is in fact obvious if one remembers that the initial data in the varying units (and thus with the time $t'$) is independent of $\epsilon $. Thus, the coefficients of the expansions: $p_0$, $p_1$, etc., are functions of ${\bf x}$ and $t'$ but they are by definition independent of $\epsilon $. With this in mind, the insertion of (\ref{matter-expansion}) into the initial data (\ref{PN_IC1})-(\ref{PN_IC3}) gives the initial data for the coefficients: at the initial time,
\be \label{PN_IC-p0p1}
p_0({\bf x}) = p^{(1)}({\bf x}), \quad p_1({\bf x}) = 0,
\ee
\be \label{PN_IC-mu0mu1}
\mu^{\ast}_0({\bf x}) = \mu^{\ast(1)}({\bf x}), \quad \mu^{\ast}_1({\bf x}) = 0,
\ee
\be \label{PN_IC-u0u1}
{\bf u} _0({\bf x}) = {\bf u} ^{(1)}({\bf x}), \quad {\bf u} _1({\bf x}) = {\bf 0}.
\ee

\subsection{Expanded field equations} \label{Expansion-field-eqs}

It is easy to check that the calculations done by Weinberg \cite{Weinberg} in his Sections 9.1 and 9.3 follow exactly from the expansions (\ref{matter-expansion}) and (\ref{metric-expansion-fixed}) of the matter fields and the metric. (We stay in fixed units with $c=1$ until the end of Section \ref{AsymptoGR}.) One difference is that, because the relevant time variable in the asymptotic expansions is the dynamical time $t'\equiv \epsilon t$, the coefficients in the expansions, e.g. $\,_1 g_{ij }$ in Eq.~(\ref{metric-expansion-fixed})$_2$, are functions of $t'$ (and of the space coordinates $x^k$). Hence, when differentiating the $\epsilon$-dependent fields with respect to $x^0 = t$, as this occurs e.g. in the definition of the connection and the Ricci tensor, one increases of one the order in $\epsilon $ (which is what Weinberg essentially assumes), but also one ends up with a derivative of the expansion coefficient with respect to $t'$. For instance, in uniform conditions, differentiating Eq.~(\ref{metric-expansion-fixed})$_2$ gives
\be \label{example-time-differentiation}
\frac{\partial g_{ij}}{\partial t} = \frac{\partial (_1g_{ij})}{\partial t'}\,\epsilon ^3 + O(\epsilon ^5).
\ee
Another difference is that the unique small parameter $\epsilon $ now appears explicitly in the expansions (but of course it does not in the expanded equations, which obtain by coefficient identification in the expansions). Thus, for instance, the components $R_{00}$, $R_{i0}$ and $R_{ij}$ of the Ricci tensor are indeed of order 2, 3 and 2 in $\epsilon$ respectively, Eqs.~(9.1.23-25) in Ref. \cite{Weinberg}, but (at least for a fluid), the corresponding components $T^{00}$, $T^{i0}$ and $T^{ij}$ of tensor ${\bf T}$ are now respectively of order 2, 3 and 4, in the same $\epsilon $, hence the expansions
\be \label{orders-T}
T^{\mu \nu} = \epsilon ^{n_{\mu \nu }} \left (_0T^{\mu \nu} +\,_1T^{\mu \nu}\,\epsilon ^2 +O(\epsilon ^4)\right ), \qquad n_{00}=2, \quad n_{i0} = 3, \quad n_{ij}=4.
\ee
\{Compare Eqs.~(9.1.42-44) of Weinberg \cite{Weinberg}.\} Defining tensor $S_{\mu \nu }\equiv T_{\mu \nu }-\frac{1}{2}g_{\mu \nu } T^\lambda _\lambda $, one then finds that $S_{00}$, $S_{i0}$ and $S_{ij}$  do have expansions with the same orders as $R_{00}$, $R_{i0}$ and $R_{ij}$ respectively, the coefficients of these expansions being given by his Eqs.~(9.1.49-52). The Einstein equations (\ref{HilbertEinstein}), equivalent to $R_{\mu \nu }=-8\pi G S_{\mu \nu }$, thus split to Weinberg's equations (9.1.53-56). It follows that the expansion (\ref{metric-expansion-fixed}) of the metric is explicitly [his Eqs.~(9.1.57-58,60-61,63-64)]:
\be \label{metric-expansion-explicit}
g_{00 } = -1 -2\Phi \epsilon ^2 -2(\Phi^2+\psi )\epsilon ^4+ O(\epsilon ^6), \quad g_{ij } =  \delta _{ij }(1-2\Phi \epsilon ^{2}) + O(\epsilon ^4), \quad g_{0i } =2\zeta _i\,\epsilon ^{3} + O(\epsilon ^5), 
\ee
the potentials $\Phi, \psi $ and $\zeta _i$ being solution of the Poisson equations
\be \label{Poisson-phi}
\Delta \Phi = 4\pi G \ _0T^{00},
\ee
\be \label{Poisson-psi}
\Delta \psi = \frac{\partial ^2 \Phi }{\partial t'^2} + 4\pi G (_1T^{00}+\ _0T^{ii}),
\ee
\be \label{Poisson-zeta}
\Delta \zeta _i = 16 \pi G \  _0T^{i0},
\ee
where $\Delta $ is the Laplacian corresponding to the Euclidean metric which has components $\delta _{ij}$ in the given harmonic coordinate system utilized, thus $\Delta \varphi \equiv \varphi_{,ii}$. \\

In the same way, starting from the general expansion (\ref{orders-T}) of the energy-momentum tensor ${\bf T}$ and the explicit expansion (\ref{metric-expansion-explicit}) of the metric $\Mat{g}$, one checks straightforwardly that the same expanded equations are derived from the local dynamical equations $T^{\mu \nu }_{;\nu } = 0$ as Eqs.~(9.3.2-5) in Ref. \cite{Weinberg}, namely
\be \label{Dynamic-expansion-time-0order}
\frac{\partial ( _0T^{00})}{\partial t'}+ \frac{\partial(_0T^{0j})}{\partial x^j}=0,
\ee
\be \label{Dynamic-expansion-space-0order}
\frac{\partial ( _0T^{0i})}{\partial t'}+ \frac{\partial(_0T^{ij})}{\partial x^j}=-\,_0T^{00}\frac{\partial\Phi }{\partial x^i},
\ee
\be \label{Dynamic-expansion-time-1order}
\frac{\partial ( _1T^{00})}{\partial t'}+ \frac{\partial(_1T^{0j})}{\partial x^j}=\,_0T^{00}\frac{\partial\Phi }{\partial t'},
\ee
\begin{eqnarray} \label{Dynamic-expansion-space-1order}
\frac{\partial ( _1T^{0i})}{\partial t'}+ \frac{\partial(_1T^{ij})}{\partial x^j} & = & -\,_1T^{00}\frac{\partial\Phi }{\partial x^i} -\,_0T^{00}\left[\frac{\partial (2\Phi^2+\psi)}{\partial x^i}+\frac{\partial \zeta_i}{\partial t'}\right] + \,_0T^{0j}\left(\frac{\partial\zeta_j }{\partial x^i}-\frac{\partial\zeta _i }{\partial x^j}\right) \nonumber\\
& & + 4 \ _0T^{0i}\,\frac{\partial\Phi }{\partial t'} + 4\ _0T^{ij}\frac{\partial\Phi }{\partial x^j} - \,_0T^{jj}\frac{\partial\Phi }{\partial x^i}.
\end{eqnarray} 

Note that, so far, the assumption of a perfect fluid was necessary only to ensure that the components of tensor ${\bf T}$ have orders in $\epsilon $ given by Eq.~(\ref{orders-T}), which can be easily checked to be consistent with the expansions (\ref{matter-expansion}) of the matter fields and with the definition of tensor ${\bf T}$ for a perfect fluid, Eq.~(\ref{T-fluid}) above. (The next Subsection shows the explicit expansions of tensor ${\bf T}$ that one gets thus.) Of course, much more general constitutive laws would still lead to the same orders for tensor ${\bf T}$. One would then have to adapt the initial data (\ref{PN_IC1})-(\ref{PN_IC3}) and the expansions (\ref{matter-expansion}) correspondingly, however.

\subsection{Explicit expansion of the energy-momentum tensor and expanded equations for a perfect fluid} \label{Explicit-expansions}

To obtain the PN field equations for a perfect fluid, we must first write the explicit expansion of the energy-momentum tensor of a perfect fluid as function of the expansions of the matter fields and the metric field. We insert the matter fields expansions (\ref{matter-expansion}) and the expansions of $g^{\mu \nu }$ and $U^0$ deduced from (\ref{metric-expansion-fixed}) by Weinberg, into the expression of the energy-momentum tensor for a perfect fluid (\ref{T-fluid}), in which $c=1$. This leads easily to the explicit expansion of the energy-momentum tensor: 
\be \label{expans-T00}
_0T^{00} = \mu^{\ast}_0, \quad _1T^{00} = \mu ^{\ast}_1 +\mu^{\ast}_0({\bf u}_0^2-2\Phi ),
\ee
\be \label{expans-T0i}
_0T^{0i} = \mu^{\ast}_0 u_0^i, \quad _1T^{0i} = [ \mu ^{\ast}_1 +\mu^{\ast}_0({\bf u}_0^2-2\Phi )+ p_0 ]u_0^i +\mu^{\ast}_0 u_1^i,
\ee
\be \label{expans-Tij}
_0T^{ij} = \mu^{\ast}_0 u_0^i u_0^j +p_0 \delta_{ij} , \quad _1T^{ij} = [\mu ^{\ast}_1+\mu^{\ast}_0({\bf u}_0^2-2\Phi )+ p_0]u_0^i u_0^j +\mu^{\ast}_0 u_1^i u_0^j + \mu^{\ast}_0 u_0^i u_1^j + p_1\delta_{ij} + 2p_0\Phi \delta_{ij}.
\ee

This expansion is different from that obtained by Weinberg \{Eqs.~(9.8.4-6), (9.8.10-13) in Ref. \cite{Weinberg}\}. This is because Weinberg does not expand the matter fields of a perfect fluid: $p$, $\mu ^{\ast}$ and ${\bf u}$. This same difference between the standard PN scheme of Fock \cite{Fock64} and Chandrasekhar \cite{Chandra65} and the asymptotic expansion method has previously been noted by Futamase and Schutz \cite{FutaSchutz} and by Rendall \cite{Rendall92}. Weinberg's approach is distinct from those of Fock and Chandrasekhar in that, in a first step, Weinberg derives the PN equations for a general form of the energy-momentum tensor, which tensor he does expand. For this reason, Weinberg's equations essentially coincide with those got with the asymptotic scheme (the differences being explained at the beginning of Subsect. \ref{Expansion-field-eqs}), until the explicit expansion of tensor ${\bf T}$ for a fluid is written. From this stage, Weinberg's equations [his Sect. (9.8)] coincide with those of Chandrasekhar \cite{Chandra65}, which are essentially equivalent to those of Fock \cite{Fock64}. In Weinberg's work, the departure from the asymptotic scheme can be seen e.g. in the fact that the same field denoted $\rho $ by Weinberg cannot be interpreted at the same time as the exact proper energy density, which we note $\mu ^{\ast}$, as it is in his Eq.~(9.8.1), and as the lowest-order coefficient in the expansion of $T^{00}$, thus our $\mu ^{\ast}_0$, as it is in his Eqs.~(9.8.4) and (9.8.9). Moreover, Eq.~(\ref{expans-T00})$_2$ here, that gives the following coefficient in the expansion of $T^{00}$ in the asymptotic scheme, has the correction term $\mu ^{\ast}_1$ as compared with the corresponding equation of Weinberg, his Eq.~(9.8.10). As suggested by Futamase and Schutz \cite{FutaSchutz}, the interpretation of the unexpanded matter fields of the standard scheme that is closest to bridging the gap with the asymptotic scheme, is to consider them as the second approximations of the exact fields. This interpretation means assuming the following correspondence between Weinberg's notation in his Eqs.~(9.8.4-15), and the present notation:
\be \label{corres}
\rho \leftrightarrow \mu ^{\ast}_{(1)} \equiv \epsilon ^2(\mu ^{\ast}_0 + \mu ^{\ast}_1 \epsilon ^2), \quad p \leftrightarrow  p_{(1)}\equiv \epsilon ^4(p_0 + p_1 \epsilon ^2), \quad {\bf v} \leftrightarrow {\bf u}_{(1)}\equiv \epsilon ( {\bf u}_0 + {\bf u}_1 \epsilon ^2).
\ee
With this interpretation, his Newtonian equations [Eqs.~(9.8.4-9)] are valid up to $O(\epsilon ^2)$ terms not included, and his 1PN equations [Eqs.~(9.8.10-15)] should be valid up to $O(\epsilon ^4)$ terms not included---provided that, when needed, one inserts the relevant power of $\epsilon $, the small parameter being incorporated in the coefficients in Weinberg's notation. Thus, for instance, Eq. (9.8.4) of Weinberg becomes
\be \label{9.8.4Weinberg}
_0T^{00} \epsilon ^2 = \mu ^{\ast}_{(1)} (1+O(\epsilon ^2)),
\ee
and his Poisson equation (9.8.9) for the Newtonian potential $\Phi $ becomes
\be \label{Poisson-phi-Weinberg}
\epsilon ^2 \Delta \Phi = 4\pi G \ \mu ^{\ast}_{(1)} (1+O(\epsilon ^2)), \quad \mathrm{\  or} \quad \Delta \Phi = 4\pi G (\mu ^{\ast}_0 + \mu ^{\ast}_1 \epsilon ^2) +O(\epsilon ^2),
\ee
which is obviously compatible with the exact equation of the asymptotic scheme, Eq.~(\ref{Poisson-phi}) with (\ref{expans-T00})$_1$:
\be \label{Poisson-phi-explicit}
\Delta \Phi = 4\pi G \mu ^{\ast}_0.
\ee
However, if one uses Eq.~(\ref{Poisson-phi-Weinberg}) instead of the exact equation (\ref{Poisson-phi-explicit}) to compute $\Phi$, then $\Phi $ is determined only up to unknown $O(\epsilon ^2)$ terms, really. (Indeed, except at the initial time, Eq.~(\ref{PN_IC-mu0mu1}), the coefficient $\mu ^{\ast}_1$ will of course not be zero.) But the Newtonian potential $\Phi $ intervenes already at the lowest order in the 1PN equation of motion [Eq. (9.8.15)], namely by the term $-\rho \nabla \Phi $ in this equation. Hence, if one does interpret Weinberg's unexpanded matter fields as the second approximations of the exact fields, thus writing Weinberg's equation (9.8.9) as Eq.~(\ref{Poisson-phi-Weinberg}) above, then his 1PN equation of motion (9.8.15) is accurate only up to unknown $O(\epsilon ^2)$ terms, in fact---i.e., it is not more accurate than the Newtonian equation of motion. This is the reason why, in our opinion, the standard PN scheme is not compatible with the asymptotic scheme. We note that the scheme used by Damour {\it et al.} for 1PN approximation \cite{DSX91,DSX92} does not pertain to the standard scheme of Fock and Chandrasekhar, because, in the DSX scheme, neither the gravitational field nor the matter fields are expanded: DSX consider the second-approximation fields, which we note ${\bf u}_{(1)}, p_{(1)},\ _{(1)}g_{\mu \nu}$, etc., and they do not split them into zero-order and first-order parts. In the standard scheme, the gravitational field is expanded (split), but the matter fields are not. In the asymptotic scheme, all fields are expanded. Hence, we feel that the DSX scheme is compatible with the asymptotic scheme, though it differs from the latter. As to Racine and Flanagan \cite{RacineFlanagan}, they use that part of Weinberg's approach \{$\S \S $ (9.1) and (9.3) in Ref. \cite{Weinberg}\} which is compatible with the asymptotic scheme. \\

Our aim in the present paper is to follow the asymptotic scheme until tractable equations of motion are got. To do that, we will moreover assume {\it barotropic} perfect fluids (one fluid per astronomical body). Recall first that the proper energy density $\mu^{\ast}$ that enters the expression (\ref{T-fluid}) of tensor ${\bf T}$ for a perfect fluid is, precisely, the sum of the proper volume densities of rest-mass and of (elastic) internal energy, $\mu^{\ast} \equiv \rho^{\ast}(1 + \Pi )$. The assumption of a barotropic fluid means that $\rho ^{\ast} = F(p)$ depend only on the pressure $p$, as well as does $\Pi $, the latter being given by \cite{Fock64}
\be \label{PiBarotropic}
				    \Pi = G(p) \equiv  \int_0 ^p \frac{\dd q}{F(q)} -\frac{p}{F(p)}.
\ee
For a barotropic fluid, it is hence convenient to replace the initial condition (\ref{PN_IC2}) for the proper energy density $\mu ^{\ast}$ by one for the proper rest-mass density $\rho ^{\ast}$: 
\begin{equation}\label{PN_IC2bis}
\mathrm{at\ } t=0,\quad \rho ^{\ast (\epsilon)}(\mathbf{x})=\epsilon^2 \rho ^{\ast (1)}(\mathbf{x})\quad \mathrm{instead\ of\quad}\mu^{\ast (\epsilon)}(\mathbf{x})=\epsilon^2 \mu^{\ast (1)}(\mathbf{x}),
\end{equation}
which makes $\rho ^{\ast} $ order 2 in $\epsilon $, like $\mu  ^{\ast }$. Accordingly, the initial condition for the expansion coefficients (\ref{PN_IC-mu0mu1}) is replaced by
\be \label{PN_IC-rho*0rho*1}
\mathrm{at\ } t=0,\quad \rho ^{\ast}_0({\bf x}) = \rho ^{\ast(1)}({\bf x}), \quad \rho ^{\ast}_1({\bf x}) = 0.
\ee
To ensure that the pressure and density fields obey Eqs.~(\ref{PN_IC1}) and (\ref{PN_IC2bis}) simultaneously, one assumes that the function defining the barotropic state equation for system S$_\epsilon $ is \cite{FutaSchutz,A23}
\be \label{StateEqn-epsilon}
F_\epsilon (p) =\epsilon ^2 F_1(\epsilon ^{ - 4} p ).
\ee
The expansions of the auxiliary matter fields $\rho ^{\ast} $, $\Pi $, and $\mu  ^{\ast} $ follow then, of course, from the expansion (\ref{matter-expansion})$_1$ of the leading field $p$ \cite{A23}:
\be \label{expans-rho*}
\rho^{\ast} = \epsilon ^2(\rho ^{\ast}_0 + \rho ^{\ast}_1 \epsilon ^2+ O(\epsilon ^4)), \quad \rho ^{\ast}_0 = F_1(p_0), \quad \rho ^{\ast}_1 = p_1 F'_1(p_0),
\ee
\be \label{expans-Pi}
\Pi = \epsilon ^2(\Pi _0 + O(\epsilon ^2)), \quad \Pi_0 = G_1(p_0), 
\ee
\be \label{expans-mu*-barotropic}
\mu ^{\ast}_0 = \rho ^{\ast}_0 , \quad \mu ^{\ast}_1 =\rho ^{\ast}_0 \Pi _0 + \rho ^{\ast}_1 .
\ee
Since Eq.~(\ref{PiBarotropic}) is equivalent to the isentropy equation: $\dd\Pi =-p\,\dd(1/\rho ^{\ast})$, a barotropic fluid is isentropic; hence, as shown by Chandrasekhar \cite{Chandra69}, its rest-mass is exactly conserved. This suggests introducing the density of rest-mass with respect to the Euclidean volume measure $\dd V \equiv \dd x^1 \dd x^2 \dd x^3$ associated with the considered harmonic coordinate system \cite{Fock64},
\be \label{rho}
\rho \equiv \sqrt{-g} \, U^0 \rho^{\ast},
\ee 
which obeys thus the usual continuity equation with the velocity ${\bf u}$. The expansion of $\rho $ is
\be \label{expans-rho}
\rho = \epsilon ^2(\rho _0 + \rho _1 \epsilon ^2+ O(\epsilon ^4)), \quad \rho _0 = \rho ^{\ast}_0 = \mu ^{\ast}_0, \quad \rho _1 = \rho ^{\ast}_1 +\rho _0(-3\Phi +{\bf u}_0^2/2).
\ee
Using (\ref{expans-rho})$_2$ in (\ref{expans-T00})$_1$, (\ref{expans-T0i})$_1$ and (\ref{expans-Tij})$_1$, and reporting in Eqs.~(\ref{Dynamic-expansion-time-0order})-(\ref{Dynamic-expansion-space-0order}), one checks that the latter ones reduce to the Newtonian equations in which $\rho _0 $ plays the role of the Newtonian density:
\begin{equation}\label{T-ord0}
  \partial _{t'} \rho_0 + \partial _j (\rho_0 u_0^j) = 0,
\end{equation}
\begin{equation}\label{i-ord0}
    \partial _{t'} (\rho_0 u_0^i) + \partial _j (\rho_0 u_0^i u_0^j)=-\rho_0
    \Phi _{,i}-p_{0,i}.
\end{equation}
Thus, Eq.~(\ref{Dynamic-expansion-time-0order}) \{Eq.~(9.3.2) in Ref. \cite{Weinberg}\} expresses the zero-order conservation of mass, not the PN mass conservation. But, rewriting the $_1T^{\mu \nu }$ 's [Eqs.~(\ref{expans-T00})$_2$-(\ref{expans-Tij})$_2$] in terms of $\rho _0$ and $\rho _1$, one gets (\ref{Dynamic-expansion-time-1order}) as 
\begin{equation}\label{T-ord1}
  \partial _{t'} (w_0+\rho_1) + \partial _j [(w_0+p_0+\rho_1)u_0^j +\rho_0
  u_1^{j}]=\rho_0\, \partial_{t'} \Phi,\qquad w_0 \equiv\rho_0\left(\frac{\mathbf{u}_0^2}{2}+\Phi +\Pi_0 \right),
\end{equation}
which, combined with the Newtonian energy equation deduced in a standard way from (\ref{T-ord0}) and (\ref{i-ord0}), gives \cite{Rendall92,A23} the order-one component of the continuity equation: 
\begin{equation}\label{mass-ord1}
  \partial _{t'} \rho_1 + \partial _j(\rho_1 u_0^j + \rho_0 u_1^j)=0.
\end{equation}
It remains to find the field equation for 1PN correction to the fluid motion. Setting
\be \label{theta_1}
\theta_1 \equiv  \rho_1 + \rho_0 \left(\frac{{\bf u}_0^2}{2} -3\Phi  + \Pi_0 \right) +p_0, 
\ee
\be\label{sigma_1}
\sigma _1 \equiv \,_1 T^{00} + \,_0 T^{jj} = \rho _1 + \rho_0\left(\frac{3}{2}{\bf u}_0^2+\Phi +\Pi_0\right)+3p_0,
\ee
we may first rewrite the expansion coefficients (\ref{expans-T0i})$_2$ and (\ref{expans-Tij})$_2$ as
\be \label{expans-T0i-theta}
_1T^{0i} = \rho _0 u_1^i + \theta _1u_0^i+4(_0T^{0i})\Phi,
\ee
\be
_1T^{ij} = \rho_0 u_0^i u_1^j  +\rho_0 u_1^i u_0^j + \theta _1 u_0^i u_0^j + (p_1-2p_0\Phi )\delta _{ij} + 4(_0T^{ij})\Phi  .
\ee
Inserting this into Eq.~(\ref{Dynamic-expansion-space-1order}), and simplifying terms with Eq.~(\ref{Dynamic-expansion-space-0order}), we obtain the sought-for equation:
\begin{eqnarray}\label{i-ord1}
    \partial _{t'} (\rho_0 u_1^i + \theta_1 u_0^i) + \partial _j ( \rho_0 u_0^i u_1^j + \rho_0 u_1^i u_0^j +\theta_1 u_0^i u_0^j)+\partial _i (p_1-2p_0\Phi )& & \nonumber \\ = -\sigma_1 \Phi _{,i} - \rho_0 [\psi _{,i}+ \partial _{t'} \zeta _{i}+(\zeta _{i,k}-\zeta _{k,i})u_0^k]. &  &
\end{eqnarray}

\section{General form of the 1PN equations of motion for a perfect-fluid system in G.R. (in the harmonic gauge)} \label{EMMC-GR-general}
\subsection{The definition of the mass centers and its motivation} \label{def-masscent}
In GR, as in any relativistic theory of gravitation, any form of material energy must both contribute to the gravitational field and be subjected to its action. It is not obvious, therefore, to state which energy density may be used as a weight function so as to define relevant mass centers. Two arguments justify the choice made \cite{A25}, of the rest-mass density  $\rho$ in the global reference frame, Eq.~(\ref{rho}): {\bf i}) since this density obeys exactly the usual continuity equation, one may commute time differentiation and barycentration.  (In other words, {\it the velocity of the mass center of the body equals the average velocity in the body.}) This was advocated by Will \cite{Will71a} as a practical advantage, which it certainly is. In our opinion, this property is also an essential feature of the classical (Newtonian) definition of the mass center, without which the notion of a mass center becomes less useful physically; for the loss of this property means that, following the motion of all constituents of the body, one could not tell what is the motion of the mass center. {\bf ii}) Rest-mass is well-correlated with astronomical observations, because it is indeed the presence of matter in the usual sense, thus characterized by its rest-mass density, that leads to the electromagnetic emission detected by the telescope. In practice, one determines an ``optical center" which, once corrected from the ``phase effects," is used to define the astronomical (observational) position of the mass center \cite{Lindegren77,Fienga99}. Hence, when giving the theoretical definition of the latter, one has indeed to check that the density selected is well-correlated with the luminous density. Thus, in our opinion, the mass centers defined with the help of $\rho$ are directly relevant to the astronomical observations. Several works have also used the choice of the rest-mass density (\ref{rho}) as the weight function to define the mass centers, e.g. Fock \cite{Fock64}, Will \cite{Will71a}, Brumberg \cite{Brumberg91}. (These works were based on the standard PNA scheme, however, which is distinct from the asymptotic scheme used in the present work: see Sect. \ref{AsymptoGR}.) In the literature on the PN equations of motion in GR, it has been more usual to define the mass centers as the local barycenters of a density $\rho '$ which is obtained by adding three (small) densities to $\rho $: the density of internal energy, the energy density associated with the self Newtonian potential of the relevant body, and the density of the kinetic energy associated with its motion with respect to a local frame attached to the mass center of that body. See e.g. Misner et al. \cite{MTW}, Spyrou \cite{Spyrou78}, Will \cite{Will93} \{cf. Eqs.~(6.21) and (6.25) in the latter work\}. Instead of the continuity equation, that density $\rho '$ would obey a balance equation with a source term (this equation is not written in the quoted works \cite{MTW,Spyrou78,Will93}). In the DSX formalism, the definition of the mass centers is more involved \{see around Eq.~(5.10) in Ref. \cite{DSX91}\}, but it also is not based on a density obeying the usual continuity equation. For sure, {\bf i}) does not hold true with such choices, although the final equations of motion may be simpler. In addition, such densities as $\rho '$ might be slightly less well correlated with the luminous density, because there is no reason that the latter increases with, for instance, the local density of the proper kinetic energy. However, the three additional densities are, of course, very small [$O(c^{-2})$] as compared with~$\rho$. More precisely, they are of the order $\rho\times GM'_a/(c^2 r_a)$ where $M'_a$ is the total mass-energy in body $(a)$ and $r_a$ is its size. Therefore, by itself, the difference in the densities $\rho $ and $\rho '$ which may be chosen as the weight function can imply only small and non-secular differences in the positions of the mass centers, at least in a weakly-gravitating system such as the solar system.\\

According to DSX \cite{DSX91}, one should consider carefully-chosen local reference frames to properly analyse the ``internal problem" of celestial mechanics (i.e., that of determining the motion of each body around its own mass center), in order to avoid introducing metrical effects due to the global velocity and the external potential. (See also Kopeikin and Vlasov \cite{KopeikinVlasov04}.) In the present work, we limit the PN calculations to those necessary to get the translational equations, thus to solve the ``external problem" in the terminology of DSX. Using the expanded field equations of the asymptotic scheme, derived in the foregoing Section, we shall derive in this Section a general form of the 1PN translational equations for the ``$\rho $-centers," Eq.~(\ref{masscent-ord1}). The remaining work to solve the external problem at the 1PN level is just to compute the integrals entering that equation. We shall compute these integrals up to and including the terms of the order 3 in the separation parameter (see Sect. \ref{GoodSeparation} for a discussion of the good separation and the way we account for it). By using the simplifying assumption of a rigid Newtonian motion for each body [Eq.~(\ref{u-rigid})], the only dynamical equations that we still have to use for that purpose are the {\it Newtonian} equations [in addition to the field equations for the gravitational potentials, Eqs.~(\ref{Poisson-phi}), (\ref{Poisson-psi}) and (\ref{Poisson-zeta})]. The Newtonian dynamical equations, when they are used, are written in the global reference frame: we do not have to introduce any ``body-attached" reference frame. \\

Thus, we define the exact masses and mass centers through the
rest-mass density $\rho$ [Eq.~(\ref{rho})]:
\begin{equation}\label{defmasscent}
  M_a \equiv\int_{\mathrm{D}_a}\rho \dd V,\qquad M_a \mathbf{a}\equiv\int_{\mathrm{D}_a}\rho\mathbf{x}\dd V
\end{equation}
where $\mathrm{D}_a$ is the (time-dependent) domain made of the spatial positions ${\bf x} \equiv (x^i)$ of the particles constituting
body $(a)$ ($a=1,..., N$) in the considered harmonic coordinate system $(x^\mu )$. At the 1PN approximation, the mass and the mass
center are approximated by
\footnote{
Henceforth, we shall use the expansions written with the ``varying units" of mass and time, $[\mathrm{M}]_\epsilon = \epsilon^2 [\mathrm{M}]$ and $\ [\mathrm{T}]_\epsilon = [\mathrm{T}]/\epsilon$, hence the expansion parameter is $\epsilon^2 = c^{-2}$. Note that the expansion coefficients and the expanded equations derived in Sect. \ref{AsymptoGR} are valid in any units (although the expansions themselves are somewhat different in fixed units and in the varying units). What we are really interested in, is the physically given system S, corresponding to the finite small value $\epsilon _0$ of $\epsilon $. Therefore, the distinction between the times $t$ and $t'=\epsilon t$ has no interest any more and we shall omit the prime.
}
\begin{equation}\label{defPNmass}
  M_a^{(1)}=M^0_a+M_a^1/c^2,\qquad M^0_a\equiv \int_{\mathrm{D}_a}\rho_0
  \dd V,\qquad M_a^1\equiv\int_{\mathrm{D}_a}\rho_1 \dd V,
\end{equation}
\begin{equation}\label{defPNmasscent}
  M_a^{(1)}\mathbf{a}_{(1)}\equiv\int_{\mathrm{D}_a}\rho_{(1)}\mathbf{x}\dd V=
  M^0_a\mathbf{a} _{0} +M_a^{1}\mathbf{a} _{1} /c^2,
\end{equation}
with
\begin{equation}\label{defmasscent-ord0-ord1}
  M^0_a\mathbf{a}_{0} \equiv \int_{\mathrm{D}_a}\rho_0\mathbf{x}\dd V,\qquad M_a^{1}\mathbf{a}_{1}\equiv \int_{\mathrm{D}_a}\rho_{1}\mathbf{x}\dd V.
\end{equation}
Note that $M^0_a$ and $\mathbf{a}_{0} $ are the Newtonian mass and mass
center. Using Eqs.~(\ref{T-ord0}) and (\ref{mass-ord1}), one shows easily \cite{A25}: {\bf i})
that $M^0_a$ and $M_a^1$ are constant in time (exactly so, insofar as the matter fields cancel on the boundaries $\partial{\mathrm{D}_a}$---in fact they are negligible there), and {\bf ii}) that 
\begin{equation}\label{commut-timediff-spaceaverage}
M_a^{(1)}\dot{\mathbf{a}}_{(1)} = \int_{\mathrm{D}_a}\rho_{(1)} \mathbf{u}_{(1)} \dd V + O(c^{-4}), \quad \mathbf{u}_{(1)} \equiv \mathbf{u}_0 + \mathbf{u}_1 \, c^{-2}, 
 \end{equation}
which is, at the 1PN level, the commuting property of time differentiation and barycentration, referred to at the beginning of this Section; in direct connection with this, we have also \cite{A25}
\be\label{M1a-apoint1}
\int_{\mathrm{D}_a}\, (\rho_1 {\bf u}_0 + \rho_0 {\bf u}_1)  \dd V = M_a^{1}\dot{\mathbf{a}}_{1} +O(c^{-2})
\ee
(this will be used later). Moreover, one finds from (\ref{defPNmass})-(\ref{defmasscent-ord0-ord1}) that the PN correction to the position of the mass center is given by
\be\label{delta-a}
{\bf a}_{(1)} - {\bf a}_0 = \frac{M_a^1}{c^2M_a^0}({\bf a}_1 - {\bf a}_0) + O(c^{-4}).
\ee
In the final equations of motion, we shall also use the notation
\begin{equation}\label{def-posi}
  \mathbf{x}_{0a} \equiv \mathbf{a}_0,
\quad \mathbf{x}_{1a} \equiv
  c^2(\mathbf{a}_{(1)}-\mathbf{a}_0),
\qquad \mathbf{x}_{a} \equiv \mathbf{a}_{(1)}= \mathbf{x}_{0a}+
\mathbf{x}_{1a}c^{-2},
\end{equation}
\begin{equation}\label{def-velo}
 \mathbf{v}_{a}\equiv \dot{\mathbf{a}}_{(1)} = \dot{\mathbf{x}}_ a = \dot{\mathbf{x}}_{0a}+
\dot{\mathbf{x}}_{1a}c^{-2}.
\end{equation}

\subsection{General form of the 1PN equations of motion}
As mentioned in the Introduction, the (1PN) equations of motion of the mass centers (EMMC's) are got by integrating the spatial components of the 1PN field equations of motion in the domains D$_a$. Since the equations for the orders zero and one are separated [Eqs.~(\ref{i-ord0}) and (\ref{i-ord1})], the same occurs for the EMMC's. The integration of the zero-order field equation gives simply the Newtonian equation of motion
\begin{equation}\label{masscent-ord0}
  M^0_a\,\ddot{a}_0^i= -\int_{\mathrm{D}_a} \rho_0 \Phi ^{(a)}_{,i}\, \dd V,
\end{equation}
where an upper dot indicates time derivative, and where we use Fock's \cite{Fock64} decomposition
\begin{eqnarray}\label{FockDecompos}
\Phi & =  & \Phi  ^{(a)} + \phi _{a}, \\
\Phi  ^{(a)}( {\bf x} ) & \equiv & -G \sum_{b \neq a} \int_{\mathrm{D}_b} \, \rho_0({\bf y}) \dd V({\bf y})/\abs{{\bf x-y} }, \\
\phi _{a}( {\bf x} ) & \equiv & -G \int_{\mathrm{D}_a} \, \rho_0({\bf y}) \dd V({\bf y})/\abs{{\bf x-y}}
\end{eqnarray}
(the corresponding usual expression of $\Phi  $ being the solution of Eq.~(\ref{Poisson-phi}) under the usual boundary condition). When integrating the field equation (\ref{i-ord1}) for the PN correction, we note that, due to Eq.~(\ref{T-ord0}), 
\be\label{K'}
\int_{\mathrm{D}_a} \, \rho_0 [\partial _{t} \zeta _{i}+\zeta _{i,k}u_0^k]\,\dd V= \frac{\dd}{\dd t} \left( \int_{\mathrm{D}_a} \,\rho_0 \zeta _{i}\,\dd V \right),
\ee
and, assuming that the matter fields cancel on the boundaries $\partial{\mathrm{D}_a}$, we  get 
\be \label{integ-i-ord1}
\frac{\dd}{\dd t} \left( \int_{\mathrm{D}_a} [\rho_0 (u_1^i +\zeta _{i})+ \theta _1 u_0^i ] \dd  V \right) = \int_{\mathrm{D}_a} f_1^i \dd V,
\ee
with
\be \label{f1i}
f_1^i = -\sigma _1 \Phi_{,i} - \rho_0\, \psi _{,i} +\rho_0\, \zeta _{k,i}\,u_0^k.
\ee
With the help of (\ref{theta_1}) and (\ref{M1a-apoint1}), this may be rewritten as
\begin{equation}\label{masscent-ord1}
  M_a^1\,\ddot{ {\bf a}}_1+\dot{{\bf I}}^{a}={\bf J}^{a}+{\bf K}^{a} +O(c^{-2}),
\end{equation}
where 
\begin{equation}\label{Iai}
  {\bf I}^{a} \equiv (I^{ai}), \quad I^{ai} \equiv \int_{\mathrm{D}_a} \left\{\left[p_0+\rho_0\left(\frac{\mathbf{u}_0^2}{2}- 3\Phi  +\Pi _0\right)\right]u_0^i+ \rho_0 \zeta_i\right\}\, \dd V,  
\end{equation}
\begin{equation}\label{Jai}
  J^{ai}\equiv \int_{\mathrm{D}_a} (-\sigma_1 \Phi  _{,i} - \rho_0 \psi_{,i})\, \dd V, 
\end{equation}
and
\begin{equation}\label{Kai}
  K^{ai}\equiv \int_{\mathrm{D}_a} \rho_0\, \zeta_{k,i}\,u_0^k \,\dd V.
\end{equation}
Owing to Eq.~(\ref{delta-a}), Eq.~(\ref{masscent-ord1}) allows to compute the 1PN correction to the acceleration of the 1PN mass centers ${\bf a}_{(1)}$:
\be\label{delta-addot}
\ddot{{\bf a}}_{(1)} - \ddot{{\bf a}}_0 = \ddot{\mathbf{x}}_{1a}c^{-2}= \frac{-\dot{{\bf I}}^{a}+{\bf J}^{a}+{\bf K}^{a}-M_a^1 \ddot{{\bf a}}_0}{c^2M_a^0}+ O(c^{-4}).
\ee
This is is the general form of the 1PN equation of motion for a perfect-fluid system in GR (in the harmonic gauge), according to the asymptotic scheme. To use this equation in practice, we must bring the integrals (\ref{Iai})-(\ref{Kai}) to a tractable form, using relevant simplifications. This is done in the next Sections. However, we can already see that these integrals all depend on the Newtonian and 1PN matter fields, hence on the internal structure of the gravitating bodies. It is hence a priori clear that, unless a ``miraculous" cancellation would occur, in (\ref{delta-addot}), of the {\it different} ways in which the internal structure influences the integrals (\ref{Iai})-(\ref{Kai}), the dependence on the internal structure should subsist in the final equations of motion.

\section{Accounting for the good separation between celestial bodies} \label{GoodSeparation}
The ``good separation" between the bodies of the system of interest means that the separation parameter \cite{A26}
\begin{equation}\label{eta}
  \eta_0\equiv\max_{a \neq b}(r_b/\abs{\mathbf{a-b}})
  \qquad (r_b\equiv \frac{1}{2}\mathrm{Sup}_{\mathbf{x},\mathbf{y}\in\mathrm{D}_b}\abs{\mathbf{x-y}})
\end{equation}
is small. We assume that the system of interest is described accurately enough by the equations of the asymptotic 1PN approximation, presented in the foregoing Sections---thus, a 1PN system S$'$ is substituted for the ``exact" system S. Then, to take the good separation into account in a clean asymptotic framework, we introduce again a (conceptual) family of systems (S$'^\eta$), each being this time a 1PN system, and with $\mathrm{S}'= \mathrm{S}'^{\eta _0}$. The family (S$'^\eta$) is defined by initial conditions \cite{A32}. We refer to Ref. \cite{A32} for a more complete motivation of this approach, involving references to the literature. As pointed out there \cite{A32}, this approach seems to be new and came from the realization that, in our previous work \cite{A26}, the lack of considering the small parameter $\eta $ within a such definite asymptotic framework (based on a conceptual family of well-separated systems) had led to the inappropriate neglect of some numerically significant terms in the 1PN equations of motion of our alternative scalar theory.  Those terms turned out to be of order $\eta ^3$ \cite{A32}. This order should be enough in the main solar system, for which we have $\eta_0 = \text{(radius of the Sun)}/\text{(minimum Mercury-Sun distance)} \simeq 1.4 \times 10^{-2}$, and $\eta_0^3 \simeq 3\times 10^{-6}$. In particular, using equations of motion derived from our alternative scalar theory and taking into account terms up to and including $\eta ^3$, we could reproduce the solar-system ephemeris DE403 of the JPL \cite{Standish95} up to a $3''/\mathrm{cy}$ difference \cite{B21}, the latter being due to the difference in the theories and in the approximation schemes. \\

Let us state the initial conditions for system S$'^\eta$. Owing to the 1PN equations, it  turns out to be sufficient to define the initial zero-order density and velocity fields. To define $\rho_0(t = 0)$, we first define the initial position of the mass centers in system
$\mathrm{S}'^\eta$:
\begin{equation}\label{a-eta-t=0}
 \mathbf{a}_0^\eta(t = 0) = \mathbf{a}_0(t = 0) \eta_0/\eta.
\end{equation}
Equation (\ref{a-eta-t=0}) ensures that, at least near $t
= 0$, the separation distances between bodies are of order
$\eta^{-1}$:
\begin{equation}\label{a-eta-b-eta}
  (r_{ab}^0)^\eta \equiv \abs{\mathbf{a}_0^{\eta}-\mathbf{b}_0^{\eta}}=
  \mathrm{ord}(\eta^{-1}) \quad \mathrm{for\ }a \neq b.
\end{equation}
Then, we just have to define the initial shape and size of each body $(a)$ in system S$'^\eta$, as some deformation of the initial shape and size of $(a)$ in the system of interest corresponding to $\eta _0$. The simplest is to assume that in fact the
bodies themselves do not depend on the separation parameter $\eta$. This is expressed by the following definition of the density $\rho_0^{\eta}(t=0)$:
\begin{equation}\label{rho-eta-t=0}
  \rho_0^{\eta}(\mathbf{x},t=0)=\rho_0(\mathbf{a+y},t=0)\quad
  \mathrm{if}\quad\mathbf{x}= \mathbf{a}_0^\eta + \mathbf{y}\quad\mathrm{with}\quad\mathbf{a}_0+{\bf y}\in \mathrm{D}_a.
\end{equation}
Equation (\ref{rho-eta-t=0}) defines the field $\rho_0^{\eta}(t=0)$ so that it is independent of $\eta$ [setting $\rho_0^{\eta}(\mathbf{x}, t = 0) = 0$ if $\mathbf{x}$ does not have
the form above for some $a = 1, ..., N$]. \\

To define the velocity $\mathbf{u}_0^\eta(t = 0)$, we use the
assumption that each body undergoes a rigid motion at
the Newtonian approximation:
\begin{equation}\label{u-rigid}
  u_0^i =  \dot{a}_0^i + \Omega^{(a)}_{ji}(x^j - a_0^j ),\ (\Omega^{(a)}_{ji} +\Omega^{(a)}_{ij} = 0),\quad \mathrm{or}\quad
  \mathbf{u}_0 =\dot{\mathbf{a}}_0+ \boldsymbol{\omega}_{a}\wedge(\mathbf{x}-{\bf a}_0),
  \quad \mathrm{for} \ \mathbf{x}\in\mathrm{D}_a.
\end{equation}
This assumption is discussed in Appendix \ref{Justif-rigid}. It is shown there that this assumption is consistent to the accuracy aimed at in this work, i.e., to get the EMMC's up to the order $\eta ^3$ included. We define the initial translation velocities of system $\mathrm{S}'^\eta$  as
\begin{equation}\label{adot-eta-t=0}
  (\dot{a}_0^i)^\eta (t=0)=(\eta/\eta_0)^{1/2}\dot{a}_0^i(t=0)
\end{equation}
and the initial spin velocities by
\begin{equation}\label{omega-eta-t=0}
  (\Omega^{(a)}_{ji})^\eta(t=0) =(\eta/\eta_0)^{1/2}\,\Omega^{(a)}_{ji}(t=0).
\end{equation}
Thus, the fields $\rho_0^\eta (t = 0)$ and $\mathbf{u}_0^\eta(t = 0)$ are well-defined, and it follows from the 1PN equations that in fact all 1PN fields are then defined at $t=0$. We assume that, as a consequence of this initial condition, the ord$(\eta ^{-1})$ separation (\ref{a-eta-b-eta}) holds true at any time in a relevant interval, in which one has moreover, consistently with (\ref{adot-eta-t=0}) and (\ref{omega-eta-t=0}),
\begin{equation}\label{adot-eta}
  (\dot{a}_0^i)^\eta = \mathrm{ord}(\eta^{1/2})
\end{equation}
and 
\begin{equation}\label{omega-eta}
  (\Omega^{(a)}_{ji})^\eta = \mathrm{ord}(\eta^{1/2}).
\end{equation}
Assumption (\ref{adot-eta}) for the magnitude of the Newtonian velocities is justified by the Newtonian estimate in a system with a dominating body, say body $(N)$:
\be \label{adot-estimate}
\dot{{\bf a}}_0^2 \approx 2 GM^0_N/r^0_{aN}
\ee
together with the good separation (\ref{a-eta-b-eta}). In the solar system, the angular velocities of the main bodies are quite small: at most of the same magnitude, in linear values, as the translation velocities---which is consistent with (\ref{adot-eta}) and (\ref{omega-eta}), accounting for the fact that the size of the bodies is $\text{ord}(\eta ^0)$. But in addition, the spins, including their axes, are very nearly constant. Since the spin evolution is determined by the rotational equations of motion, we cannot simply assume that the rates $\dot{\Omega}^{(a)}_{ji}$ are very small, but have to derive it from these equations. In the asymptotic scheme used here, the zero-order (Newtonian) equations apply exactly to the zero-order fields, hence the the rates $\dot{\Omega}^{(a)}_{ji}$ are determined by the Newtonian rotational equations. We shall prove in Appendix \ref{SpinEvolution} that these rates are $O(\eta ^3)$, provided that: {\bf i}) the system is made of rigidly-rotating well-separated bodies, in the sense of Eqs.~(\ref{a-eta-b-eta}), (\ref{u-rigid}), (\ref{adot-eta}) and (\ref{omega-eta}), {\it and} \ {\bf ii}) these bodies have a ``quasi-spherical" inertia tensor, in the sense that
\be \label{quasi-spheric}
\abs{\gamma ^{(a)}_i - \gamma ^{(a)}_k} = O(\eta^2) \quad (a = 1,..., N;\ i,k=1,2,3),
\ee
where the $\gamma ^{(a)}_i$ 's are the eigenvalues of the inertia tensor $I^{(a)}_{ik}$ defined in Eq.~(\ref{def-Iaij-Omegaa}). This amounts approximately (or exactly, for homogeneous bodies) to assuming that the difference between the linear dimensions of each body in the different directions is $O(\eta)$. It would be easy to modify the definition of the initial density field (\ref{rho-eta-t=0}) so that it becomes compatible with (\ref{quasi-spheric}), e.g. by introducing an orthogonal affinity centered at ${\bf a}_0$ and leaving one dimension of the body unchanged while adjusting the other two. Thus, to account in an asymptotic framework for the numerical situation in the solar system, we need that our conceptual family of 1PN systems couples the good separation with the quasisphericity.
\\

In Appendix \ref{Integrals-eta}, we use the orders in $\eta $ given by (\ref{a-eta-b-eta}), (\ref{adot-eta}), (\ref{omega-eta}) to evaluate the integrals (\ref{Iai})-(\ref{Kai}) up to and including ord$(\eta ^3)$ terms. [Of course, the density fields $\rho_0$, $\rho _1$, etc., as well as the masses $M^0_a$ and $M^1_a$ and other energy integrals, are ord$(\eta^0)$.] There, we use some of the calculations done in Refs. \cite{A26,A32}, and also we use Newtonian or purely mathematical calculations of Fock \cite{Fock64}. We shall use the quasi-sphericity assumption (\ref{quasi-spheric}), leading to the estimate (\ref{Omega-dot=O(eta^3)}), only in two cases: to get Eq.~(\ref{Iai-dot-explicit}) giving $\dot{I}^{ai}$, and to get Eq.~(\ref{Lai-self-dot}).

\section{Explicit 1PN equations of motion for well-separated, rigidly-rotating, quasi-spherical bodies}\label{EMMC-GR-explicit}

The explicit form of the 1PN correction to the equations of motion is got by inserting the explicit form of the integrals $\dot{\bf {I}}^{a}, {\bf {J}}^{a}, {\bf {K}}^{a}$, Eqs.~(\ref{Iai-dot-explicit}), (\ref{Ja-La-M1addota}), (\ref{Lai-self-dot}), (\ref{Lai-external-explicit}), and (\ref{Kai-explicit}), into the general form (\ref{delta-addot}) of the 1PN correction. In a first step, we obtain thus
\begin{eqnarray} \label{EMMC-GRH-separated}
\ddot{\mathbf{x}}_{1a} & = &  \left[\frac{-\dot{{\bf a}}_0^2}{2} +3 \Phi^{(a)}({\bf a}_0) -\frac{ 6T_a +2\varepsilon _a}{3M_a^0} \right] \ddot{{\bf a}}_0   
+ G \sum_{b \neq a}  M_b^0 \frac{ \left[{\bf n}^0_{ab} .(4 \dot{{\bf a}}_0 - 3\dot{{\bf b}}_0)\right] (\dot{{\bf a}}_0 -\dot{{\bf b}}_0 )}{(r^0_{ab})^2} \nonumber  \\
 & & - G\sum_{b\neq a} \left\{ M^1_b +M^0_b\left[2(\dot{{\bf a}}_0-\dot{{\bf b}}_0)^2 -\frac{\dot{{\bf a}}_0^2}{2} -\frac{3}{2} ({\bf n}^0_{ab} .\dot{{\bf b}}_0) ^2+ \Phi^{(a)}({\bf a}_0) +\Phi^{(b)}({\bf b}_0) \right] + \varepsilon _{ab}\right\} \frac{ {\bf n}^0_{ab} } {(r^0_{ab})^2} \nonumber\\
&  & + G\sum_{b\neq a} M^0_b \left\{ \frac{\mathbf{x}_{1b}\mathbf{-}\mathbf{x}_{1a}+
  3\left[(\mathbf{x}_{1a}\mathbf{-}\mathbf{x}_{1b})\mathbf{.}\mathbf{n}^0_{ab}
  \right]\mathbf{n}^0_{ab}} {(r^0_{ab})^3} +\frac{({\bf n}^0_{ab} .\ddot{{\bf b}}_0){\bf n}^0_{ab} +7\ddot{{\bf b}}_0} {2r^0_{ab}} \right\} + \frac{{\bf M}^{(a)}.\ddot{{\bf a}}_0}{M^0_a} \nonumber\\
&  & - \frac{\dot{{\bf I}} ^{a}_\text{ns}}{M^0_a} + {\bf j}^{\,a}_\text{ns} +\frac{ {\bf L}^{a}_{1\,\text{ns}} +{\bf L}^{a}_{2\,\text{ns}} }{M_a^0} +O(\eta^{7/2})+O(c^{-2}) , \\
\varepsilon _{ab} & \equiv & 8\left[T_a\frac{M_b^0}{M_a^0}+T_b\right] + \frac{2}{3}\left[\varepsilon_a\frac{M_b^0}{M_a^0}+\varepsilon _b\right], 
\end{eqnarray}
in which the space tensor ${\bf M}^{(a)}$, related to the spin ${\bf \Omega}^{(a)}$, is given by Eq.~(\ref{def-Mmatrix}), and the (spatial) vectors $\dot{{\bf I}} ^{a}_\text{ns}$, ${\bf j}^{\,a}_\text{ns}$, ${\bf L}^{a}_{1\,\text{ns}}$  and ${\bf L}^{a}_{2\,\text{ns}}$ are given by Eqs.~(\ref{Iai-ns}), (\ref{Ja-La-M1addota-ns}), (\ref{L1ai-ns}) and (\ref{L2ai-ns}), respectively. The four latter quantities reduce to zero if the Newtonian density fields are assumed spherical in the sense of Eq.~(\ref{rho-spheric}). It may be worth to repeat here that, in the asymptotic scheme which has been used in this work, the equations for the zero-order quantities and the 1PN corrections are separated \cite{FutaSchutz,Rendall92,A23} (see Sect. \ref{AsymptoGR}). For this reason, it is not directly possible to compare Eq.~(\ref{EMMC-GRH-separated}) with the LD equation. However, if we define the vector radius in terms of the full 1PN positions (\ref{def-posi}):
\be\label{def-nab-rab}
r_{ab} \equiv \abs{ {\bf x}_{a} - {\bf x}_{b} }, \qquad \mathbf{n}_{ab} \equiv \frac{{\bf x}_a - {\bf x}_b}{r_{ab}},
\ee
instead of defining it in terms of the zero-order positions, as in (\ref{def-n0ab-r0ab}), then it is easy to derive the following 1PN expansion:
\be\label{relation-nab-n0ab}
\frac{\mathbf{n}_{ab} } {r_{ab}^2} = \frac{\mathbf{n}^0_{ab} }{(r^0_{ab})^2} +\frac{1}{c^2} \left\{\frac{ \mathbf{x}_{1a} -\mathbf{x}_{1b} -3\left[(\mathbf{x}_{1a}\mathbf{-}\mathbf{x}_{1b})\mathbf{.}\mathbf{n}^0_{ab}
  \right]\mathbf{n}^0_{ab}}{r_{ab}^3}  \right\} +O\left(c^{-4}\right).
\ee
We use this and the fact that, up to $O(\eta ^4)$ terms, one may use the spherical estimate (\ref{spherical-phi-ext}) for the external Newtonian potentials $\Phi^{(a)}({\bf a}_0)$ and $\Phi^{(b)}({\bf b}_0)$ in Eq.~(\ref{EMMC-GRH-separated}). [This is independent of the ``optional" sphericity assumption (\ref{rho-spheric}). Moreover, we use the expression (\ref{M1a-explicit}) for $M^1_b$ in~(\ref{EMMC-GRH-separated}) and we switch to the notation (\ref{def-posi})-(\ref{def-velo}).] This enables us to group Eq.~(\ref{EMMC-GRH-separated}) with the Newtonian equation of motion (\ref{masscent-ord0}) so as to get
\begin{eqnarray} \label{EMMC-GRH-grouped}
\ddot{\mathbf{x}}_{a} & = &  -\sum_{b\neq a} \frac{ G M_b^0}{r_{ab}^2} {\bf n}_{ab} \left\{ 1+ \delta_b + \frac{1}{c^2} \left[ {\bf v}_a^2 + 2 {\bf v}_b^2 -4{\bf v}_a.{\bf v}_b -\frac{3}{2}( {\bf n}_{ab} .{\bf v}_b)^2 -4\sum_{d\neq a} \frac{ G M_d^0}{r_{ad}} \right.\right. \nonumber\\
& & \left.\left.-\sum_{d\neq b} \frac{ G M_d^0}{r_{bd}}\left(1+\frac{r_{ab}}{2r_{db}} {\bf n}_{ab}.{\bf n}_{db}\right)  + \frac{6T_a}{M^0_a} \right] \right\} -\frac{7}{2} \sum_{b\neq a} \frac{ G M_b^0}{r_{ab}} \sum_{d\neq b} \frac{ G M_d^0}{c^2 r_{bd}} {\bf n}_{bd} \nonumber\\ 
& & + \sum_{b\neq a} \frac{ G M_b^0}{c^2 r_{ab}^2} \left\{\left[{\bf n}_{ab} .(4 {\bf v}_a - 3{\bf v}_b)\right] ({\bf v}_a -{\bf v}_b )- \frac{{\bf M}^{(a)}.{\bf n}_{ab}}{M^0_a} \right\}\nonumber\\                                                          
& & +\frac{1}{c^2} \left[ - \frac{\dot{{\bf I}} ^{a}_\text{ns}}{M^0_a} + {\bf j}^{\,a}_\text{ns} + \frac{ {\bf L}^{a}_{1\,\text{ns}} + {\bf L}^{a}_{2\,\text{ns}} }{M_a^0} \right] +O(\eta^{7/2})+O(c^{-4}), 
\end{eqnarray}
with, as before, matrix ${\bf M}^{(a)}$ of Eq.~(\ref{def-Mmatrix}), and with
\be
\delta_b \equiv \frac{1}{c^2} \left[ 3\sum_{d\neq b} \frac{ G M_d^0}{r_{bd}(t=0)}+\frac{1}{2}{\bf v}_b^2(t=0) +\frac{27T_b+20\varepsilon_b}{3M^0_b} \right] \ll 1.
\ee
The $\delta_b$ 's are constant up to $O(\eta^{7/2})$, see Eq.~(\ref{Ta-dot}). Obviously, the term in~(\ref{EMMC-GRH-grouped}) that involves this small constant may be absorbed, up to $O(c^{-4})$ corrections, in a redefinition of the zero-order masses, replacing them in~(\ref{EMMC-GRH-grouped}) by
\be
M'^0_b \equiv M^0_b(1+\delta_b).
\ee
Moreover, we recall that $\dot{{\bf I}} ^{a}_\text{ns}$, ${\bf j}^{\,a}_\text{ns}$, ${\bf L}^{a}_{1\,\text{ns}}$ and $ {\bf L}^{a}_{2\,\text{ns}} $ are given by Eqs.~(\ref{Iai-ns}), (\ref{Ja-La-M1addota-ns}), (\ref{L1ai-ns}) and (\ref{L2ai-ns}), respectively, and that these terms cancel in the ``spherical" case in the sense of Eq.~(\ref{rho-spheric}). In that ``spherical" case, the r.h.s. of~(\ref{EMMC-GRH-grouped}) coincides exactly with the LD acceleration ${\bf A}_a^\text{LD}$, apart from a term depending on the spin and on the internal structure:
\begin{eqnarray}\label{deltaA}
\ddot{\mathbf{x}}_{a} - {\bf A}_a^\text{LD} & = & \left(\frac{6T_a}{M^0_a}+\frac{{\bf M}^{(a)}}{M^0_a}\right).\left(-\sum_{b\neq a} \frac{ G M_b^0}{c^2r_{ab}^2} {\bf n}_{ab}\right) =\left(\frac{6T_a}{M^0_a}+\frac{{\bf M}^{(a)}}{M^0_a}\right).\frac{\ddot{\mathbf{x}}_{0a}}{c^2} +O(\eta ^4) + O(c^{-4})\nonumber\\
& = & \frac{\gamma _a}{c^2M^0_a} \left[ 6(\omega^{(a)})^2 + {\bf \Omega}^{(a)}.{\bf \Omega}^{(a)} \right].\ddot{{\bf x}}_{0a}+O(\eta ^4) + O(c^{-4}).
\end{eqnarray}
Here, $\gamma _a$ is the spherical inertia moment (\ref{Iik-spheric}), which does depend on the internal structure, and $(\omega^{(a)})^2 \equiv \Omega^{(a)}_{jk} \Omega^{(a)}_{jk}/2$ is the square of the angular velocity. \{${\bf \Omega}^{(a)}.{\bf \Omega}^{(a)}$ is the ``tensor" [see the footnote with Eq.~(\ref{def-Mmatrix})] with $ik$ component $\Omega^{(a)}_{ij} \Omega^{(a)}_{jk}$.\} We note that this new term is order 3 in $\eta$. It is clearly different from the two spin-orbit terms and the spin-spin term appearing in the monopole-dipole equations of DSX \cite{DSX92}, e.g. because, on the basis of Eqs.~(\ref{a-eta-b-eta}), (\ref{adot-eta}) and (\ref{omega-eta}) here, the latter terms \{Eqs.~(6.32,33,34) in Ref. \cite{DSX92}\} are of order $\eta ^4$, $\eta ^4$ and $\eta ^5$, respectively. Many more structure-dependent parameters appear in the non-spherical case, see Eqs.~(\ref{Ja-La-M1addota-ns}), (\ref{L1ai-ns}) and (\ref{L2ai-ns}). But, as discussed in the Introduction, the departure from sphericity is unlikely to give numerically important 1PN corrections. 

\section{Summary and conclusion} \label{Conclusion}

We started from the family of initial data proposed by Futamase \& Schutz \cite{FutaSchutz}, which defines a family $(\mathrm{S}_\epsilon)$ of perfectly-fluid gravitating systems. This family is obtained by applying a Newtonian exact similarity transformation to the initial data defining the matter fields for the system of interest, $\mathrm{S}=\mathrm{S}_{\epsilon_0}$, and by assuming for the spatial metric an initial data which enforces spatial isotropy. That family allows to define the PN approximation in a clean asymptotic framework \cite{FutaSchutz}. (A similar family, differing only in the initial data for the gravitational field, allows to do the same for an alternative scalar theory \cite{A23}.) We call the corresponding PN scheme ``the asymptotic scheme." By using a change of units depending on the parameter $\epsilon$ \cite{A23}, we are naturally led to postulate definite expansions for the independent fields, Eq.~(\ref{matter-expansion}) for the matter fields and Eq.~(\ref{metric-expansion-fixed}) for the metric. The latter is equivalent to the expansion of Weinberg \cite{Weinberg}, but Weinberg did not expand the matter fields. However, in a first step, he did expand the energy-momentum tensor ${\bf T}$, by assuming, for its different components, expansions in terms of the typical velocity in the system, his Eqs.~(9.1.42-44). This is a slightly different expansion from the one we postulated in terms of the unique small parameter $\epsilon$, Eq.~(\ref{orders-T}) here, but, together with the expansion of the metric, it leads to the same expanded equations: the Einstein equations, written in terms of the 1PN gravitational potentials, lead to Eqs.~(\ref{Poisson-phi})-(\ref{Poisson-zeta}), while the dynamical equation leads to Eqs.~(\ref{Dynamic-expansion-time-0order})-(\ref{Dynamic-expansion-space-1order}) for tensor ${\bf T}$. But since we do expand the matter fields, the explicit dynamical equations, as written in terms of the matter fields, are definitely different from those of Weinberg \cite{Weinberg} in his Sect. 9.8, which coincide with those obtained by Chandrasekhar \cite{Chandra65}. The hydrodynamical equation for the 1PN corrections according to the asymptotic scheme valid for GR in the harmonic gauge was written here explicitly for the first time, Eq.~(\ref{i-ord1}): Futamase and Schutz \{\cite{FutaSchutz}, Eq.~(4.28)$_2$ with (4.27)$_{2-3}$ \}, as well as Rendall \{\cite{Rendall92}, Eq.~(44)\}, did not write this equation in a such fully-explicit form. Moreover, in these two works, equations for the ``$\frac{1}{2}$PN" order were considered necessary. In the present work, we used $\epsilon $-dependent units in which the small parameter becomes $\epsilon = c^{-1}$ and all fields are order zero (thus justifying the formal expansions in powers of $c^{-1}$) and in which, moreover, only $\epsilon^2 = c^{-2}$ plays a role in most equations---thus suggesting to postulate Taylor expansions in $c^{-2}$, for which there is no $\frac{1}{2}$PN term. In fixed units, this leads to an expansion of the metric and a general expansion of the energy-momentum tensor that are equivalent to those postulated by Weinberg \cite{Weinberg}, involving merely even powers of $\epsilon $ or merely odd powers. The calculations of his Sects. 9.1 and 9.3 show that these expansions are consistent.\\

We used the asymptotic PN scheme to derive 1PN equations of motion for the mass centers (EMMC's) in a weakly-self-gravitating system, according to GR (in the harmonic gauge). This was not done (with the asymptotic scheme) before. Like Fock \cite{Fock64}, Will \cite{Will71a}, and Brumberg \cite{Brumberg91}, we define the mass centers through the rest-mass density, Eq.~(\ref{defmasscent}), and we obtain the EMMC's by integrating the 1PN dynamical equations in the domain occupied by the relevant body. One difference is that we use the asymptotic PN scheme instead of the ``standard" PN scheme proposed by Fock \cite{Fock64} and by Chandrasekhar \cite{Chandra65}. As a consequence, we get separated EMMC's for the zeroth (Newtonian) order and for the first PN correction. In a first step, these are general equations valid for any 1PN system, Eqs.~(\ref{masscent-ord0}) and (\ref{delta-addot}). These general equations show that one has a priori to expect an influence of the structure of the gravitating body (e.g. the density profile) and its internal motion (e.g. the rotation velocity), since they depend on integrals of all matter fields. Another important difference with previous works (not only with Refs. \cite{Fock64}, \cite{Will71a}, and \cite{Brumberg91}), is that we use a definite asymptotic framework for the separation parameter $\eta$, see Sect. \ref{GoodSeparation}, which enables us to obtain results including an asymptotic error estimate. In the present work, we got explicitly the 1PN correction to the EMMC's when including the terms up to the order $\eta ^3$ included, Eq.~(\ref{EMMC-GRH-separated}). To do that, we assumed, merely when calculating the 1PN {\it correction}, that, at the {\it Newtonian} approximation, the bodies are rigidly rotating (as is nearly the case for the main bodies of the solar system), and quasi-spherical in the sense of Eq.~(\ref{quasi-spheric}) (which simulates in an asymptotic framework the real situation for these same bodies). \\

It turns out that the separated EMMC's of the zeroth and first order can be grouped together to give Eq.~(\ref{EMMC-GRH-grouped}). The latter may be readily compared with the Lorentz-Droste (Einstein-Infeld-Hoffmann) equation. In addition to new (structure-dependent) 1PN terms accounting for the departure from exact spherical symmetry, there is a new 1PN term depending on the internal structure and on the rotation velocity of the body considered, Eq.~(\ref{deltaA}). It is worth to emphasize that the present derivation depends essentially on the weak-field assumption and is invalid for a strong-field system [although the definition of the mass center through the rest-mass density, Eqs.~(\ref{rho}) and ~(\ref{defmasscent}), makes sense also for a strong field]. Thus, for a binary system of two
neutron stars, each maximally spinning, the new term might lead to
10 \% corrections to Newtonian physics, whereas in fact, if the binary is widely
separated, the Newtonian dynamics is known to be a good approximation.

\section*{Acknowledgment}
I am grateful to the referee for his remarks: i) on the smallness of the influence of the different densities which may rightly be selected to define the mass centers, and ii) on the fact that the absolute magnitude of the perturbation to the acceleration may be misleading; and also, for providing the example of a binary system of neutron stars, which confirms that the present results do not apply to a strong-field system.

\appendix
\section{The integrals ${\bf I}^{a}$, ${\bf J}^{a}$ and ${\bf K}^{a}$ for well-separated, rigidly-rotating bodies} \label{Integrals-eta}

Let us first note the following correspondence between the notations of Weinberg \cite{Weinberg}, used in the present work, and those of Fock, also used in Refs. \cite{A25,A26,A32} \{see Eqs.~(\ref{Poisson-zeta}), (\ref{expans-T0i}) and (\ref{expans-rho}) here, and Eq.~(73.03) in Fock \cite{Fock64}\}:
\be\label{corres}
\quad  \Phi \leftrightarrow  -U, \quad \zeta _i \leftrightarrow -4 U_i.\\
\ee

\subsection{Integral ${\bf I}^{a}$}

This integral [Eq.~(\ref{Iai})] is equal to
\begin{eqnarray} \label{Iai-IaiETGv1}
I^{ai} & = & \int_{\mathrm{D}_a} \left[p_0+\rho_0\left(\frac{\mathbf{u}_0^2}{2}-\Phi  +\Pi _0\right)\right]u_0^i \, \dd V \nonumber \\
& & + 2 \int_{\mathrm{D}_a} (-\rho_0 \Phi^{(a)} u_0^i) \, \dd V + 2 \int_{\mathrm{D}_a} (-\rho_0\phi_a u_0^i) \, \dd V + \int_{\mathrm{D}_a} \rho_0 \zeta_i \, \dd V\nonumber \\ 
& \equiv & I^{ai}_1 + 2 I^{ai}_2 + 2 I^{ai}_3 +I^{ai} _4.   
\end{eqnarray}
Integral $I^{ai}_1$ on the r.h.s. was in fact denoted by $I^{ai}$ in Refs. \cite{A25,A26,A32} [it indeed played the same role there as the integral (\ref{Iai}) plays in Eq.~(\ref{masscent-ord1}) here] and it is given in two pieces by Eqs.~(A20)-(A21) in Ref. \cite{A26}; the second piece, Eq.~(A21) in Ref. \cite{A26}, is just integral $I^{ai}_2$ above. In Ref. \cite{A26}, it could not be assigned a definite order in $\eta $ to the remainders, but now it is in general easy to do that, using Eqs.~(\ref{a-eta-b-eta}), (\ref{adot-eta}) and (\ref{omega-eta}).
\footnote{
An exception is for the order in Eq.~(\ref{I1ai}) here. It was proved in Ref. \cite{A33}, Eq.~(A12), that Fock's ``Lyapunov's equation" \{\cite{Fock64}, Eq.~(73.26)\} indeed applies to a body in a well-separated system, with a definite remainder in $\eta$:
\be \label{FockLyapunov}
p_0 + \rho_0 \Pi_0 = \rho_0(\Omega_a - \phi_a )+ O(\eta^3), 
\ee
where $\Omega_a$ is defined in Eq.~(\ref{def-Iaij-Omegaa}) below. [In this proof, the fact was used that $\dot{\Omega}^{(a)}_{ji}= O(\eta^3)$ as a consequence of the Newtonian rotational equations---this is proved in Appendix \ref{SpinEvolution} below, yet that proof needs quasi-spherical bodies in the sense of Eq.~(\ref{quasi-spheric}).] We have $u_0^i = \text{ord}(\eta^{1/2})$ [Eqs.~(\ref{adot-eta})-(\ref{omega-eta})]. Equation (\ref{I1ai}), including the order of the remainder, follows by a straightforward computation. 
}
We have thus:
\be \label{I1ai}
I^{ai}_1 = I^{ai}_2 + (M^0_a \dot{{\bf a}}_0^2/2 + 2 T_a + 4 \varepsilon_a) \dot{a}_0^i + (\dot{a}_0^k \Omega^{(a)}_{lk}I^{(a)}_{jl} + 2T_{aj}+ 4 \varepsilon_{aj})\Omega^{(a)}_{ji} +O(\eta ^{7/2}),
\ee
\be\label{I2ai}
I^{ai}_2 \equiv \int_{\mathrm{D}_a} (-\rho_0 \Phi^{(a)} u_0^i) \, \dd V = -M^0_a \dot{a}_0^i \Phi^{(a)}({\bf a}_0) - I^{(a)}_{jk} \Omega^{(a)}_{ki} \Phi^{(a)}_{,j}({\bf a}_0) + O(\eta ^{7/2}),
\ee
where \cite{Fock64}
\begin{equation}\label{def-epsa-epsai}
  \varepsilon_a\equiv-\int_{\mathrm{D}_a} \rho_0 \phi _a \dd V/2, \quad \varepsilon_{aj} \equiv-\int_{\mathrm{D}_a} \rho_0 \phi _a (x^j-a_0^j) \dd V/2, 
\end{equation}
\begin{equation}\label{def-Iaij-Omegaa}  
I^{(a)}_{ij}\equiv \int_{\mathrm{D}_a} \rho_0(x^i-a_0^i)(x^j-a_0^j) \dd V, \quad \Omega_a \equiv \Omega^{(a)}_{ik} \Omega^{(a)}_{jk}(x^i-a_0^i)(x^j-a_0^j)/2,
\ee
\begin{equation}\label{def-Ta-Taj}
T_a\equiv \int_{\mathrm{D}_a} \rho_0 \Omega_a \dd V,\quad T_{aj} \equiv  \int_{\mathrm{D}_a} \rho_0 \Omega_a (x^j-a_0^j) \dd V. 
\end{equation}
The third integral is immediately computable from Eq.~(\ref{u-rigid}) above, in terms of the potential-energy integrals (\ref{def-epsa-epsai}):
\be\label{I3ai}
I^{ai}_3 \equiv \int_{\mathrm{D}_a} (-\rho_0\phi_a u_0^i) \, \dd V = 2 (\varepsilon_a \dot{a}_0^i + \varepsilon_{aj}\Omega^{(a)}_{ji}).
\ee
By dividing the potential $\zeta _i$ (the negative Newtonian potential associated with the density $4\,_0T^{i0} = 4\rho_0 u_0^i$) into ``external" and ``self" parts $Z_i ^{(a)}$ and $\zeta_{ai}$, exactly as for $\Phi $ in Eq.~(\ref{FockDecompos}), one finds [cf. Fock's integrals (75.36) and (76.29-30)]:
\be\label{Iai4}
I^{ai}_4 = -4 I^{ai}_3 + I^{ai}_5,
\ee
with
\begin{eqnarray} \label{Fock76.29}
I^{ai}_5 & \equiv & \int_{\mathrm{D}_a} \rho_0 Z_i ^{(a)} \dd V  =  M_a^0 Z_i ^{(a)}({\bf a}) + O(\eta ^{7/2}) \nonumber \\
& = & -4 G M_a^0 \sum_{b \neq a} \frac{M_b^0 \dot{b}_0^i } {\abs{{\bf a}_0-{\bf b}_0}} + 4 G M_a^0 \sum_{b \neq a}   \Omega^{(b)}_{ji} I^{(b)}_{jk} \frac{\partial} {\partial a_0^k } \frac{1} {\abs{{\bf a}_0-{\bf b}_0}} + O(\eta ^{7/2}).
\end{eqnarray} 
Summing these different contributions, we get
\begin{eqnarray} \label{Iai-explicit}
I^{ai} & = & (I^{ai}_1-I^{ai}_2) + 3 I^{ai}_2 -2 I^{ai}_3 + I^{ai}_5 \nonumber \\
& = & (M^0_a \dot{{\bf a}}_0^2/2 + 2 T_a) \dot{a}_0^i + (\dot{a}_0^k \Omega^{(a)}_{lk} I^{(a)}_{jl} + 2T_{aj})\Omega^{(a)}_{ji} \nonumber \\ & & -3M^0_a \dot{a}_0^i \Phi^{(a)}({\bf a}_0)  -4 G M_a^0 \sum_{b \neq a} \frac{M_b^0 \dot{b}_0^i } {\abs{{\bf a}_0-{\bf b}_0}} \nonumber \\
& & - 3I^{(a)}_{jk} \Omega^{(a)}_{ki} \Phi^{(a)}_{,j}({\bf a}_0)+ 4 G M_a^0 \sum_{b \neq a}   \Omega^{(b)}_{ji} I^{(b)}_{jk} \frac{\partial} {\partial a_0^k } \frac{1} {\abs{{\bf a}_0-{\bf b}_0}} + O(\eta ^{7/2}).
\end{eqnarray}
When differentiating this with respect to the time, as is needed for insertion in the equation of motion (\ref{masscent-ord1}), we note that, for well-separated, rigidly-rotating and quasi-spherical bodies (at the Newtonian approximation), the rate of the (Newtonian) angular rotation velocity is $O(\eta^3)$ [Eq.~(\ref{Omega-dot=O(eta^3)})]. Moreover, independently of the quasi-sphericity assumption (\ref{quasi-spheric}), we have by the Newtonian rotational equations (\ref{rhs-angular-momentum-rate}) \{cf. Fock \cite{Fock64}, Eq.~(72.32)\}
\be\label{Ta-dot}
\frac{ \dd T_a}{\dd t} = O\left(\eta^ 3\abs{\Omega^{(a)}_{ji} }\right) =O(\eta ^{7/2}). 
\ee 
In addition, due to the assumed rigid zero-order motion (\ref{u-rigid}), the (Newtonian) inertia tensor is constant in the frame following the (Newtonian) rotation of the body, so that its time derivative in the starting harmonic coordinates is known explicitly as \cite{Fock64}
\be \label{Iik-dot}
\dot{I}^{(a)}_{ik} = \Omega^{(a)}_{ji} I^{(a)}_{jk} +\Omega^{(a)}_{jk} I^{(a)}_{ij}. 
\ee
In the same way, we have by (\ref{Omega-dot=O(eta^3)}) and Fock's Eqs.~(72.24) and (74.06)
\be\label{Tai-dot} 
\frac{ \dd T_{aj}}{\dd t} = \frac{1}{2}\Omega^{(a)}_{kj} \Omega^{(a)}_{lj} ( \Omega^{(a)}_{mi} I^{(a)}_{mkl} +\Omega^{(a)}_{mk} I^{(a)}_{iml} +\Omega^{(a)}_{ml} I^{(a)}_{ikm} )+O(\eta ^3), \quad I^{(a)}_{ikl}\equiv \int_{\mathrm{D}_a} \rho_0(x^i-a_0^i)(x^k-a_0^k)(x^l-a_0^l)  \dd V.
\ee
Therefore, we get
\begin{eqnarray} \label{Iai-dot-explicit}
\dot{{\bf I}}^{a} & = & M^0_a (\dot{{\bf a}}_0.\ddot{{\bf a}}_0)\dot{{\bf a}}_0 + \{M^0_a [\dot{{\bf a}}_0^2/2 -3\Phi^{(a)}({\bf a}_0)]+ 2 T_a\} \ddot{{\bf a}}_0 -3M^0_a \dot{{\bf a}}_0 \frac{\dd}{\dd t} [\Phi^{(a)}({\bf a}_0)]  \nonumber  \\
 & & -4 G M_a^0 \frac{\dd}{\dd t} \sum_{b \neq a} \frac{M_b^0 \dot{{\bf b}}_0 } {\abs{{\bf a}_0-{\bf b}_0}} - {\bf M}^{(a)}.\ddot{{\bf a}}_0 + \dot{{\bf I}} ^{a}_\text{ns} + O(\eta ^{7/2}),
\end{eqnarray}
\be\label{Iai-ns}
\dot{I}^{ai}_\text{ns} \equiv  \dot{a}_0^k \Omega^{(a)}_{lk} \dot{I}^{(a)}_{jl} \Omega^{(a)}_{ji} - 3\dot{I}^{(a)}_{jk} \Omega^{(a)}_{ki} \Phi^{(a)}_{,j}({\bf a}_0)+ 4 G M_a^0 \sum_{b \neq a}   \Omega^{(b)}_{ji} \dot{I}^{(b)}_{jk} \frac{\partial} {\partial a_0^k } \frac{1} {\abs{{\bf a}_0-{\bf b}_0}}+2\frac{ \dd T_{aj}}{\dd t}\Omega^{(a)}_{ji} ,
\ee
in which the rates of the inertia tensor are given by (\ref{Iik-dot}), and where
\footnote{
Recall that we are using a fixed system of harmonic coordinates $x^\mu$, such that the initial condition has the form (\ref{PN_IC1})-(\ref{ICmetric-F&S}). However, the equations written here are covariant under a ``Cartesian" spatial coordinate change, $x'^0=x^0,\  x'^i=R_{ij}x^j$ with ${\bf R} = (R_{ij})$ a (constant) orthogonal matrix, ${\bf R} \in \sf{O}(3,\sf{R})$. Hence the upper or lower indices need not be distinguished (they may be exchanged by the Euclidean metric that has components $\delta _{ij}$ in all these coordinate systems). Moreover, these equations may be brought to a form covariant under an arbitrary spatial coordinate change: $y^0=x^0, \ y^i=\psi ^i(x^j)$, provided some care is taken, e.g., in handling the time derivatives: the components of the velocity vector ${\bf v}$ along a trajectory $t\mapsto {\bf x}(t)$ are, in any space coordinates $y^i$, the time-derivatives $v^i=\dot{y}^i$, but the acceleration ${\bf A}$ must be defined as the ``absolute" time derivative of the velocity vector using the connection associated with the Euclidean metric; the components of ${\bf A}$ are $\ddot{x}'^i$ only in ``Cartesian" coordinates $x'^i$.
} 
\be\label{def-Mmatrix}
{\bf M} ^{(a)} .\ddot{{\bf a}}_0 \equiv  (M^{(a)} _{il}\ddot{ a}_0^l), \qquad  M^{(a)} _{il} \equiv \Omega^{(a)}_{ij} I^{(a)}_{jk} \Omega^{(a)}_{kl}.
\ee 
To have $\dot{{\bf I}}^{a}$ up to the $O(\eta^{7/2})$ remainder, we may use in (\ref{Iai-dot-explicit}) and (\ref{Iai-ns}) the spherical (or monopole) estimates of the Newtonian potentials; in fact the dipoles cancel due to Eq.~(\ref{defmasscent-ord0-ord1})$_1$, so that
\be \label{spherical-phi-ext}
\Phi^{(a)}({\bf a}_0) = -\sum_{b \neq a} \frac{G M_b^0 } {\abs{{\bf a}_0-{\bf b}_0}} + O(\eta ^3), \quad \frac{\dd}{\dd t}[\Phi^{(a)}({\bf a}_0)] = -\frac{\dd}{\dd t} \sum_{b \neq a} \frac{G M_b^0 } {\abs{{\bf a}_0-{\bf b}_0}} + O(\eta ^{7/2}). 
\ee
In the case that the Newtonian mass density $\rho_0$ is assumed spherical inside each body:
\begin{equation}\label{rho-spheric}
  \forall \mathbf{x}\in\mathrm{D}_a,\,\rho_0(\mathbf{x})= \rho_a(r),
  \quad r\equiv\abs{\mathbf{x-a}_0}\quad (a = 1, ..., N),
\end{equation}
we have 
\be \label{Iik-spheric}
I^{(a)}_{ik} = \gamma _a \delta _{ik}, \quad \gamma_a \equiv \frac{4\pi}{3}\int_0^{r_a} \, r^4 \rho_a(r) \dd r.
\ee
Then the rates (\ref{Iik-dot}) cancel. The same is true for the $T_{aj}$'s \cite{Fock64}, hence $\dot{I}^{ai}_\text{ns}$ cancels.

\subsection{Integral ${\bf J}^{a}$}

To calculate integral $J^{ai}$ [Eq.~(\ref{Jai})], we introduce Fock's auxiliary potential
\begin{equation}\label{W}
  W(\mathbf{x},t) \equiv \int G \abs{\mathbf{x - y}} \rho(\mathbf{y},t)\, \dd V(\mathbf{y})/2
\end{equation}
and we note that, due to Eqs.~(\ref{Poisson-psi}) and (\ref{sigma_1}), and since $\Delta W = -\Phi $, the potential $\psi $ is given by
\be \label{psi}
  -\psi = B +  \partial  ^2 W/ \partial t^2,
\end{equation}
where $B$ satisfies the Poisson equation
\be \label{Poisson-B}
\Delta B = -4\pi G \sigma_1.
\ee
Therefore, integral $J^{ai}$ is exactly the integral denoted so in Ref. \cite{A25}, Eq.~(4.11) and after, though here the field $\sigma_1$ is defined by Eq.~(\ref{sigma_1}). This integral has been studied in detail in Refs. \cite{A25}-\cite{A32}, under the assumption that the boundary condition for $B $ ensures that is is indeed the (positive) Newtonian potential associated with the matter field $\sigma_1$, the latter having spatially-compact support. We have \{\cite{A32}, Eq.~(A14)\}
\begin{eqnarray}\label{Jai-Lai}
J^{ai} - L^{ai} & = & GM^0_a \frac{\partial} {\partial x^k }_{{\bf x}={\bf a}_0} \sum_{b \neq a} \left[ \frac{\alpha_b} {\abs{{\bf x-b}_0}} + \beta_{bj} \frac{x^j-b_0^j} {\abs{{\bf x-b}_0}^3}\right] \nonumber \\
& & - \alpha _a \frac{\partial \Phi^{(a)}} {\partial a_0^i } - \beta_{aj} \frac{\partial^2 \Phi^{(a)}} {\partial a_0^i \partial a_0^j} + O(\eta ^4),
\end{eqnarray}
with
\be\label{def-Lai}
L^{ai} \equiv  \int_{\mathrm{D}_a} \rho_0 \frac{\partial^3 W} {\partial x^i \partial t^2} \, \dd V,
\ee
\be\label{def-alpha-a-beta-aj}  
\alpha_a \equiv \int_{\mathrm{D}_a} \sigma _1 \, \dd V, \quad \beta_{aj} \equiv \int_{\mathrm{D}_a} \sigma_1({\bf x}) (x^j - a_0^j) \, \dd V ({\bf x}).
\ee
The coefficients $\alpha _a$ and $\beta_{aj}$ ($a=1,...,N,\quad j=1,2,3$) defined by (\ref{def-alpha-a-beta-aj}) have been computed in Ref.~\cite{A32}, though with a slightly different $\sigma_1$ from the one valid for GR (in the harmonic gauge), Eq.~(\ref{sigma_1}) here. The calculation is easy, using the rigid velocity field (\ref{u-rigid}) together with Eq.~(\ref{FockLyapunov}) and a Taylor expansion of $\Phi^{(a)}$ at ${\bf a}_0$. Due to the presence in~(\ref{sigma_1}) of the field $p_0$, we have now to know the following integrals [Fock's Eqs.~(74.24-25)], which also follow from ``Lyapunov's equation"~(\ref{FockLyapunov}), and which have the same remainder as the latter:
\be\label{Integral-p}
3 \int_{\mathrm{D}_a} p_0 \, \dd V = \varepsilon_a - 2 T_a + O(\eta ^3),
\ee
\be\label{Integral-p(xj-aj)}
2\int_{\mathrm{D}_a} p_0(x^j-a_0^j) \, \dd V = \eta_{aj} - T_{aj} + O(\eta ^3), \quad \eta_{aj} \equiv (B^{(a)}_{k\,jk}+B^{(a)}_{j\,kk})/2,  
\ee
where \cite{Fock64}
\be \label{Bajik}
B^{(a)}_{j\,ik} \equiv \int_{\text{space}} [\delta_{ik}(\nabla \phi_a)^2/2-\phi_{a,i}\phi_{a,k}](x^j-a_0^j) \,\dd V/4\pi G.\, 
\footnote{
In this integral, the integrand, say $f$, is not Lebesgue-integrable, but we have $f=f_\text{spher} + \delta f$, where $f_\text{spher}$ is the integrand corresponding to spherical symmetry (with $\phi _a({\bf x}) = \phi _a(r), \quad r\equiv \abs{{\bf x-a}_0}$, and $\phi _a(r) = -GM_a^0/r$ for $r\geq r_a$), and where $\delta f$ is Lebesgue-integrable. Hence, the integral exists in the sense of $\int_{\text{space}} f \,\dd V \equiv \lim_{R\rightarrow \infty } \int_{\abs{{\bf x-a}_0} \leq R} f\,\dd V $, because $\int_{\abs{{\bf x-a}_0} \leq R} f_\text{spher} \,\dd V =0$ for any $R\geq 0$.
}
\ee
We find thus:
\be\label{alpha-M1a}
\alpha_a = M^1_a +M^0_a \left[\frac{3}{2} \dot{{\bf a}}_0^2 + \Phi^{(a)}({\bf a}_0) \right] + 8T_a+\frac{2}{3}\varepsilon _a +O(\eta ^3) 
\ee
with
\be\label{M1a-explicit}
M^1_a = M^0_a \left[\frac{\dot{{\bf a}}_{00}^2}{2}-3\Phi^{(a)}({\bf a}_{00}) \right] + T_a+6\varepsilon _a +O(\eta ^3), 
\ee
\be\nonumber
{\bf a}_{00} \equiv {\bf a}_0(t=0), \ \dot{{\bf a}}_{00} \equiv \dot{{\bf a}}_0(t=0), 
\ee
or
\be\label{alpha-a}
\alpha_a = M^0_a \left[\frac{3 \dot{{\bf a}}_0^2 +\dot{{\bf a}}_{00}^2}{2}+ \Phi^{(a)}({\bf a}_0) -3\Phi^{(a)}({\bf a}_{00}) \right] + 9T_a+\frac{20}{3}\varepsilon _a +O(\eta ^3), 
\ee
and
\be\label{beta-aj}
\beta_{aj} = M^1_a (a_1^j-a_0^j) + 3\dot{a}_0^k \Omega^{(a)}_{lk} I^{(a)}_{jl} + 3T_{aj} + \eta_{aj} +O(\eta^2) = M^1_a (a_1^j-a_0^j) + \eta_{aj} +O(\eta). 
\ee
To the purpose of getting the EMMC's up to $\eta^3$ included, the second estimate of $\beta_{aj}$, up to $O(\eta)$ not included, is enough since $\beta_{aj}$ is multiplied by $O(\eta^3)$ in (\ref{Jai-Lai}). To the same purpose, we may use the spherical estimate (\ref{spherical-phi-ext})$_1$ for $\Phi^{(a)}$ in the expression (\ref{alpha-a}) for $\alpha_a$, and also in the expression (\ref{masscent-ord0}) of the Newtonian acceleration $\ddot{{\bf a}}_0$, which enters the PN correction to the acceleration (\ref{delta-addot}). We also use Eqs.~(\ref{delta-a}) and (\ref{def-posi}) to reexpress the terms with the $\beta $ 's. This gives
\begin{eqnarray} \label{Ja-La-M1addota}
\frac{ {\bf J}^{a}-{\bf L}^{a}-M_a^1 \ddot{{\bf a}}_0 } {M^0_a} & = & G\sum_{b\neq a} \left[\alpha_b +(\alpha_a-M^1_a)\frac{M^0_b}{M^0_a} \right]\left(\frac{-{\bf n}^0_{ab}}{(r^0_{ab})^2}\right) + {\bf j}^{\,a}_\text{ns} \nonumber\\
&  & + G\sum_{b\neq a} \frac{M^0_b}{(r^0_{ab})^3}\left[\mathbf{x}_{1b}\mathbf{-}\mathbf{x}_{1a}+
  3\left((\mathbf{x}_{1a}\mathbf{-}\mathbf{x}_{1b})\mathbf{.}\mathbf{n}^0_{ab}
  \right)\mathbf{n}^0_{ab}\right] +O(\eta^4). 
\end{eqnarray}
Here,
\be\label{def-n0ab-r0ab}
r^0_{ab} \equiv \abs{{\bf a}_0 - {\bf b}_0} \equiv \abs{ {\bf x}_{0a} - {\bf x}_{0b} }, \qquad \mathbf{n}^0_{ab} \equiv \frac{{\bf a}_0 - {\bf b}_0}{r^0_{ab}}
\ee
and
\be \label{Ja-La-M1addota-ns}
j^{\,ai}_\text{ns} \equiv G\sum_{b\neq a} \left( \eta_{bj} - \frac{M^0_b}{M^0_a} \eta_{aj} \right) \frac{\delta _{ij}-3 n^{0i}_{ab} n^{0j}_{ab}}{(r^0_{ab})^3}.
\ee
The latter cancels, as do the $\eta_{bj}$ 's, when all bodies are spherical in the sense of (\ref{rho-spheric})---see Eqs.~(\ref{Integral-p(xj-aj)})$_2$ and (\ref{epsai-Baik-Bajik-spheric}).
\\

Integral (\ref{def-Lai}) was calculated in Ref. \cite{A26} to an order which turns out to be sufficient. It is given by \{\cite{A26}, Eqs.~(A14), (A20)\}:
\be \label{Lai-ext-self}
L^{ai} = \frac{ \dd}{\dd t} \int_{\mathrm{D}_a} \rho_0 \frac{\partial^2 w_a} {\partial x^i \partial t}\, \dd V + \int_{\mathrm{D}_a} \rho_0 \frac{\partial^3 W^{(a)}} {\partial x^i \partial t^2}\, \dd V +O(c^{-2}) \equiv L^{ai}_1 + L^{ai}_2 + O(c^{-2}),
\ee
where the ``self" part is given by the following equation \{\cite{A26}, Eq.~(A21)\}, which is an exact one:
\be \label{Lai-self}
L^{ai}_1 = \frac{ \dd}{\dd t} \left(-\varepsilon_a \dot{a}_0^i + B^{(a)}_{ik}\dot{a}_0^k - \varepsilon_{aj} \Omega^{(a)}_{ji} + B^{(a)}_{j\,ik}\Omega^{(a)}_{jk}\right),
\ee
with \cite{Fock64}
\be \label{Baik}
B^{(a)}_{ik} \equiv \int_{\text{space}} [\delta_{ik}(\nabla \phi_a)^2/2-\phi_{a,i}\phi_{a,k}] \,\dd V/4\pi G.
\ee
Due to the assumed rigid motion (\ref{u-rigid}), the rates of the quantities $\varepsilon_a, \varepsilon_{aj}, B^{(a)}_{ik}, B^{(a)}_{j\,ik}$ are known: $\dot{\varepsilon}_a=0$, $B^{(a)}_{ik}$ follows the rule (\ref{Iik-dot}), and we have \cite{Fock64}
\be \label{epsaj-dot,Bajik-dot}
\dot{\varepsilon}_{aj} = \Omega^{(a)}_{kj} \varepsilon_{ak}, \quad \dot{B}^{(a)}_{j\,ik} = \Omega^{(a)}_{lj} B^{(a)}_{l\,ik} + \Omega^{(a)}_{li} B^{(a)}_{j\,lk} + \Omega^{(a)}_{lk} B^{(a)}_{j\,il}.
\ee
Moreover, in the case that the Newtonian density $\rho _0$ is exactly spherical in body $(a)$, then we have \cite{A26} 
\be \label{epsai-Baik-Bajik-spheric}
\varepsilon_{aj} =0, \quad B^{(a)}_{ik} = \delta_{ik} \varepsilon_a/3, \quad B^{(a)}_{j\,ik} = 0.
\ee 
Recall that we shall obtain the EMMC's for a family of well-separated and quasi-spherical bodies, i.e. a family of bodies, indexed by the separation parameter $\eta$, and whose family all bodies become closer and closer to being spherical as $\eta \rightarrow 0$. For a such family, we should have
\footnote{
This does not strictly follow from the assumption (\ref{quasi-spheric}), which concerns merely the departure from sphericity of the inertia tensor. However, if one would explicitly define, in a natural way, the density field $\rho_0^{\eta }$ satisfying (\ref{quasi-spheric}), as outlined after that equation, then he should indeed obtain (\ref{epsai-Bajik-quasi-spheric}), and even probably the $\varepsilon_{aj}$ 's and the $B^{(a)}_{j\,ik}$ 's should be $O(\eta)$ or higher. To be logically consistent, we may content ourselves by simply assuming that the family does satisfy (\ref{epsai-Bajik-quasi-spheric}) in addition to (\ref{quasi-spheric}).
}
\be \label{epsai-Bajik-quasi-spheric}
\varepsilon_{aj} \rightarrow 0\quad \text{and}\quad B^{(a)}_{j\,ik} \rightarrow 0 \quad \text{as}\quad \eta \rightarrow 0.
\ee
Therefore, we get from (\ref{Lai-self}) and (\ref{Omega-dot=O(eta^3)}):
\be \label{Lai-self-dot}
{\bf L}^{a}_1 =  -\frac{2}{3}\varepsilon_a \ddot{{\bf a}}_0 + {\bf L}^{a}_{1\,\text{ns}}
\ee
with
\be\label{L1ai-ns}
L^{ai}_{1\,\text{ns}} = -\frac{1}{3}\varepsilon_a \ddot{a}_0^i + B^{(a)}_{ik}\ddot{a}_0^k + \dot{B}^{(a)}_{ik}\dot{a}_0^k - \dot{\varepsilon}_{aj} \Omega^{(a)}_{ji} + \dot{B}^{(a)}_{j\,ik}\Omega^{(a)}_{jk} +o(\eta ^{3}).
\ee
We have $L^{ai}_{1\,\text{ns}} = 0$ (exactly) if the Newtonian density fields are spherical in the sense of Eq.~(\ref{rho-spheric}).
\newline

The ``external" part is given by Eq.~(A24) of Ref. \cite{A26}, of which the last term is $O(\eta ^4)$,
\footnote{
because the assumed rigid motion implies that $\dot{I}^{(a)}_{jk}=O(\Omega_{ij})=O(\eta ^{1/2})$, see Eq.~(\ref{Iik-dot}).
}
hence may be omitted here:
\be \label{Lai-external}
L^{ai}_2 = -\frac{GM^0_a}{2} \sum_{b \neq a} M^0_b \left( \ddot{b}_0^k \frac{\partial^2 \abs{{\bf a}_0-{\bf b}_0}}  {\partial a_0^i \partial a_0^k} - \dot{b}_0^k \dot{b}_0^j \frac{\partial^3 \abs{{\bf a}_0-{\bf b}_0}}  {\partial a_0^i \partial a_0^k \partial a_0^j} \right)  + L^{ai}_{2\,\text{ns}} +O(\eta^4),
\ee
\be\label{L2ai-ns}
L^{ai}_{2\,\text{ns}} \equiv -\frac{1}{2} \sum_{b \neq a} \ddot{I}^{(b)}_{jk} \frac{\partial^3 \abs{{\bf a}_0-{\bf b}_0}}  {\partial a_0^i \partial a_0^k \partial a_0^j}
\ee
(the latter cancels in the case of spherical symmetry), or more explicitly
\be\label{Lai-external-explicit}
{\bf L}^{a}_2 = \frac{GM^0_a}{2}\sum_{b\neq a} \ M^0_b \left\{ \frac{\left(\mathbf{n}^0_{ab}\mathbf{.}\ddot{\mathbf{b}}_0\right)\mathbf{n}^0_{ab}
  -\ddot{\mathbf{b}}_0}{r^0_{ab}} 
  +\frac{\left[3(\mathbf{n}^0_{ab}\mathbf{.}\dot{\mathbf{b}}_0)^2-
  \dot{\mathbf{b}}_0^2\right]\mathbf{n}^0_{ab}
  -2\left(\mathbf{n}^0_{ab}\mathbf{.}\dot{\mathbf{b}}_0\right)\dot{\mathbf{b}}_0 } {(r^0_{ab})^2}\right \}+ {\bf L}^{a}_{2\,\text{ns}} +O(\eta^4).
\ee

\subsection{Integral ${\bf K}^{a}$}

To calculate integral $K^{ai}$ [Eq.~(\ref{Kai})], we first note that, due to the Poisson equation (\ref{Poisson-zeta}) with the expansion (\ref{expans-T0i})$_1$, we have \{Eq.~(75.27) of Fock \cite{Fock64}\}:
\be\label{Kai-self=0}
\int_{\mathrm{D}_a} \rho_0\,\zeta_{ka,i} \, u_0^k \,\dd V = 0.
\ee
Hence it merely remains the ``external" part. I.e., we have
\be\label{Kai=Kai-external}
K^{ai} = \int_{\mathrm{D}_a} \rho_0\,Z_{k,i}^{(a)} \, u_0^k \,\dd V.
\ee
To compute it, we use Fock's Eq.~(76.23), which we rewrite in the present notation, also giving the (easily-evaluated) order of the remainder:
\be\label{zeta-i-external}
Z_k ^{(a)}({\bf x}) = -4 G \sum_{b \neq a} \frac{M_b^0 \dot{b}_0^k } {\abs{{\bf x}-{\bf b}_0}} + 4 G \sum_{b \neq a}   \Omega^{(b)}_{lk} I^{(b)}_{lj} \frac{\partial} {\partial x^j } \frac{1} {\abs{{\bf x}-{\bf b}_0}} + O(\eta ^{7/2}).\\
\ee
Inserting (\ref{zeta-i-external}) and the Taylor expansion of $1/\abs{{\bf x}-{\bf b}_0}$ into Eq.~(\ref{Kai=Kai-external}) with the rigid velocity field (\ref{u-rigid}), we get
\be\label{Kai-explicit}
{\bf K}^{a} = 4 G M_a^0 \sum_{b \neq a} M_b^0 \dot{{\bf a}}_0 {\bf .} \dot{{\bf b}}_0 \frac{{\bf n}^0_{ab}}{(r^0_{ab})^2}+ O(\eta ^{4}).\\
\ee

\section{Newtonian spin evolution for a system of well-separated, rigidly-rotating, quasi-spherical bodies} \label{SpinEvolution}

We start from Fock's Eqs.~(72.06) and (72.09), which, combined together, write 
\be \label{angular-momentum-rate}
\frac{\dd}{\dd t} M^{(a)}_{ik} = \int_{\text{D}_a} \ \rho_0\left[-(x^i-a_0^i)\frac{\partial \Phi ^{(a)}}{\partial x^k} + (x^k-a_0^k)\frac{\partial \Phi ^{(a)}}{\partial x^i}\right] \dd V,
\ee
where
\be \label{angular-momentum-definition}
M^{(a)}_{ik} \equiv \int_{\text{D}_a} \ \rho_0 \left[(x^i-a_0^i) u_0^k - (x^k-a_0^k) u_0^i \right] \dd V.
\ee
Equation (\ref{angular-momentum-rate}) may be seen as the application to body $(a)$ of the Newtonian theorem stating that the rate of change of the total angular momentum of a system is equal to the sum of the external torques on the system. Recall that the Newtonian equations apply exactly to the zero-order quantities (Subsect. \ref{Explicit-expansions}), and that we are actually considering a family (S$'^\eta$) of 1PN gravitating systems (Sect. \ref{GoodSeparation}) (although we omit the superscript $\eta$ on the fields for the simplicity of notation). Due to the good separation, the r.h.s. of (\ref{angular-momentum-rate}) may be evaluated as \{Fock \cite{Fock64}, Eq.~(72.13) in which we evaluate the order of the remainder with the help of Eq.~(\ref{a-eta-b-eta}) here\}:
\be \label{rhs-angular-momentum-rate}
\frac{\dd}{\dd t} M^{(a)}_{ik} = \sum_{b \neq a} \frac{3GM_b^0 (a_0^j-b_0^j) } {\abs{{\bf a}_0-{\bf b}_0}^5} \left[(a_0^k-b_0^k) I^{(a)}_{ij} - (a_0^i-b_0^i) I^{(a)}_{kj}  \right] +O(\eta^4). 
\ee
As to the angular momentum tensor (\ref{angular-momentum-definition}), using the rigid velocity field (\ref{u-rigid}) as well as the definitions of the Newtonian mass center (\ref{defmasscent-ord0-ord1})$_1$ and the inertia tensor (\ref{def-Iaij-Omegaa})$_1$, it is easily calculated to be
\be \label{angular-momentum-rigid}
M^{(a)}_{ik} = \Omega^{(a)}_{jk} I^{(a)}_{ji}  - \Omega^{(a)}_{ji} I^{(a)}_{jk}. 
\ee
We differentiate this with respect to the time, using the rate (\ref{Iik-dot}) of the inertia tensor. {\it Then}, we rewrite the result in space coordinates $(x'^i)$ which are Cartesian for the Euclidean space metric (that one having components $\delta _{ij}$ in the given harmonic coordinate system $(x^\mu)$ utilized) and which, at the {\it current} time $t$, bring the inertia tensor of body $(a)$ to the diagonal form (the $x'^i$ 's are deduced from the $x^i$ 's by a space rotation):
\be \label{Iik-diagonal}
I^{(a)}_{ik} = \delta_{ij} \gamma ^{(a)}_j \delta_{jk}.
\ee
This yields
\be \label{lhs-angular-momentum-rate}
\frac{\dd}{\dd t} M^{(a)}_{ik} = \dot{\Omega}^{(a)}_{ik} \left( \gamma ^{(a)}_i +\gamma ^{(a)}_k\right) + \Omega^{(a)}_{ij} \Omega^{(a)}_{jk} \left(\gamma ^{(a)}_i -\gamma ^{(a)}_k\right) .
\ee
We also simplify (\ref{rhs-angular-momentum-rate}) with the help of (\ref{Iik-diagonal}), and we equate the result with (\ref{lhs-angular-momentum-rate}). We thus get in the coordinates $(x'^i)$: 
\be \label{Omega-dot}
\dot{\Omega}^{(a)}_{ik} =\frac{\gamma ^{(a)}_i -\gamma ^{(a)}_k  }{\gamma ^{(a)}_i +\gamma ^{(a)}_k} \left[\sum_{b \neq a} \frac{3G M_b^0 (a_0^i-b_0^i)(a_0^k-b_0^k) } {\abs{{\bf a}_0-{\bf b}_0}^5} - \Omega^{(a)}_{ij} \Omega^{(a)}_{jk} \right] +O(\eta^4). 
\ee
If, now, we account for the fact that $\Omega^{(a)}_{ik} = O(\eta^{1/2})$ [Eq.~(\ref{omega-eta})] and for the quasi-sphericity assumption (\ref{quasi-spheric}), we do see that 
\be \label{Omega-dot=O(eta^3)}
\dot{\Omega}^{(a)}_{ik} = O(\eta^3). 
\ee

\section{Justification of assuming a rigid rotation for well-separated bodies} \label{Justif-rigid}
Let us shortly discuss the possibility for a perfectly-fluid body in a weakly self-gravitating system to have a rigid motion at the zero-order approximation [Eq.~(\ref{u-rigid})]. Using the continuity equation (\ref{T-ord0}), we rewrite the zero-order dynamical equation (\ref{i-ord0}) as Euler's equation, and in the latter we insert the rigid velocity field~(\ref{u-rigid}). This yields
\be\label{Euler-rigid}
\rho_0 \left[\ddot{a}_0^i + \left(\dot{\Omega}_{ji} ^{(a)} - \Omega_{ik}^{(a)}\Omega_{jk}^{(a)}\right)(x^j-a_0^j)\right] = - \rho_0 \Phi _{,i}  -p_{0,i}.
\ee
Thus, the r.h.s. must depend linearly on the position ${\bf x}$, as does the l.h.s.. According to Eq.~(\ref{FockDecompos}), the Newtonian potential $\Phi$ decomposes into the self-potential $\phi_a$ and the external potential $\Phi^{(a)}$. If the body is isolated, the latter cancels. In that case, Eq.~(\ref{Euler-rigid}) can certainly be verified, for it is well-known that an isolated rotating mass made of a perfect-fluid body is dynamically possible (may be under certain restrictions on the state equation). Thus,
\be\label{Euler-rigid-isolated}
\rho_0 \left( \dot{\Omega}_{ji} ^{(a)} - \Omega_{ik}^{(a)}\Omega_{jk}^{(a)}\right)(x^j-a_0^j)= - \rho_0 \phi _{a,i}  -p_{0,i}
\ee
can be satisfied exactly. (The mass center has no acceleration in that case, of course.) The presence of external bodies produces time-varying tidal forces which prevent the exact validity of the equilibrium~(\ref{Euler-rigid}). This means that in fact the body will undergo some time-dependent deformation: tides. If we consider a family of well-separated systems, we have
\be\label{Taylor-Phiext-eta3}
\Phi^{(a)} _{,i}({\bf x}) = \Phi ^{(a)} _{,i}({\bf a}_0)  + \Phi^{(a)}_{,i,j}({\bf a}_0)(x^j-a_0^j) +O(\eta ^4) \qquad ({\bf x} \in \text{D}_a).
\ee
Inserting this Taylor expansion into the equation of motion for the zero-order mass center, Eq.~(\ref{masscent-ord0}), and accounting for (\ref{defmasscent-ord0-ord1}), we find that
\be\label{ddot-a-eta3}
\ddot{a}_0^i = \Phi^{(a)}_{,i}({\bf a}_0) +O(\eta ^4).
\ee
We can then rewrite~(\ref{Euler-rigid}) as
\be\label{Euler-rigid-eta3}
\rho_0 \left( \dot{\Omega}_{ji} ^{(a)} - \Omega_{ik}^{(a)}\Omega_{jk}^{(a)}\right)(x^j-a_0^j) = - \rho_0 \left[\phi _{a,i} + \Phi^{(a)}_{,i,j}({\bf a}_0)(x^j-a_0^j)\right] -p_{0,i} +O(\eta ^4).
\ee
This shows that, to accommodate a given spin $({\Omega}_{ji}^{(a)} )$, we cannot start from a solution of Eq.~(\ref{Euler-rigid-isolated}) and modify just the spin rate $\dot{\Omega}_{ji} ^{(a)}$ [which, we recall, is primarily subject to Eq.~(\ref{Omega-dot})]: we must also modify the pressure field $p_0$, which (through the state equation) determines the field $\rho _0$, the latter determining in turn the self-field $\phi_a$. This would need a detailed study. But if in the equilibrium equation~(\ref{Euler-rigid}) we neglect not only $O(\eta ^4)$ but even $O(\eta ^3)$, then we are left with 
\be\label{Euler-rigid-eta2}
\rho_0 \left( \dot{\Omega}_{ji} ^{(a)} - \Omega_{ik}^{(a)}\Omega_{jk}^{(a)}\right)(x^j-a_0^j) = - \rho_0 \phi _{a,i} -p_{0,i} +O(\eta ^3),
\ee
in which, owing to Eq.~(\ref{Omega-dot=O(eta^3)}), we may further neglect $\dot{\Omega}_{ji} ^{(a)}$---provided the system is quasi-spherical in the sense of Eq.~(\ref{quasi-spheric}). This is all what we used. In particular, this is what was used in Ref. \cite{A33} to derive ``Lyapunov's equation" reproduced above, Eq.~(\ref{FockLyapunov}). Thus, to get the translational equations of motion up to $\eta^3$ included, we need only to solve the Newtonian internal equilibrium up to the order $\eta^2$ included. The latter is not influenced by the external bodies, Eq.~(\ref{Euler-rigid-eta2}), hence it is compatible with a rigid rotation (at the Newtonian approximation), as postulated in Eq.~(\ref{u-rigid}). 


\end{document}